\documentclass[a4paper]{article}

\usepackage[utf8]{inputenc}

\usepackage{fullpage}

\usepackage[hyperfootnotes=false]{hyperref}

\usepackage{fancyhdr}

\usepackage{subfigure} 

\usepackage{amsmath}
\usepackage{amssymb}
\usepackage{latexsym}
\usepackage{sgame}
\usepackage{pstcol}

\usepackage{booktabs}
\usepackage{multirow}

\usepackage{lscape}

\usepackage{boxedminipage}

\usepackage{algorithm}
\usepackage{algorithmic}

\usepackage{listings}

\usepackage{endnotes}

\usepackage{minitoc}

\usepackage{ifpdf}

\usepackage{natbib}
\bibpunct{(}{)}{;}{a}{,}{}

\usepackage[parfill]{} 
\usepackage{setspace}
\usepackage{fullpage}

\ifpdf
\usepackage[pdftex]{graphicx}
\else
\usepackage{graphicx}
\fi

\title{Modeling Network Evolution Using Graph Motifs}
\author{Drew Conway}

\date{\today}

\begin{document}

\ifpdf
\DeclareGraphicsExtensions{.pdf, .jpg, .tif}
\else
\DeclareGraphicsExtensions{.eps, .jpg}
\fi

\maketitle

\abstract{Network structures are extremely important to the study of political science. Much of the data in its subfields are naturally represented as networks.  This includes trade, diplomatic and conflict relationships. The social structure of several organization is also of interest to many researchers, such as the affiliations of legislators or the relationships among terrorist.  A key aspect of studying social networks is understanding the evolutionary dynamics and the mechanism by which these structures grow and change over time.  While current methods are well suited to describe static features of networks, they are less capable of specifying models of change and simulating network evolution.  In the following paper I present a new method for modeling network growth and evolution.  This method relies on graph motifs to generate simulated network data with particular structural characteristic. This technique departs notably from current methods both in form and function.  Rather than a closed-form model, or stochastic implementation from a single class of graphs, the proposed ``graph motif model'' provides a framework for building flexible and complex models of network evolution.  The method is computationally based, relying on graph theoretic and machine learning techniques to grow networks.  The paper proceeds as follows: first a brief review of the current literature on network modeling is provided to place the graph motif model in context. Next, the graph motif model is introduced, and a simple example is provided.  As a proof of concept, three classic random graph models are recovered using the graph motif modeling method: the Erd\H{o}s-R\`{e}nyi binomial random graph, the Watts-Strogatz ``small world'' model, and the Barab\'{a}si-Albert preferential attachment model.  In the final section I discuss the results of these simulations and subsequent advantage and disadvantages presented by using this technique to model social networks.}

\newpage
\setstretch{1.9} 

The study of networks is one of the most interdisciplinary fields in contemporary scholarship.  With its origins in graph theory, and migration to the social science largely via sociology \citep{Freeman_2004}, many political scientists have discovered the value of these methods.  The primary reason for this is that much of the data relevant to political science can be represented as a network.  In network science the primary unit of analysis is the edge, or link between two actors.  Likewise, many subfields in political science study interactions and organizations that are naturally modeled as a network.  At the macro-level in international relations this includes trade, diplomatic and conflict relationships, while at  micro-level networks can be used to study the structure of terrorist organizations.  For comparative politics this may include government coalitions networks, or party affiliations.  Finally, in American politics this can include campaign finance contributions or legislative co-sponsorship networks.

Given the breadth of possible applications, network analysis is a growing methodological subfield within political science.  The work within this niche can be crudely divided into two applications of network analysis: structurally descriptive, or networks as dependent variables.  Research in the former category has a relatively long and rich history, with well-established methods for describing the structure of networks. \footnote{The most complete reference on statistically descriptive methods is \emph{Social Network Analysis: Methods and Applications}, by Wasserman and Foust \citep{wasserman_social_1994}.}  These methods include measures of actor centrality, whereby the relative position or role of actors is based on their number and type of edges within the network.  In political science these methods have most frequently been applied to international relations studies.  For example, structurally descriptive methods have been used to illustrate how social capital transferred through inter-governmental organizations' membership networks can create conflicts between states \citep{Hafner_Burton_2006}.  Centrality-based methods have also been used to identify key actors in the conflict in Chechnya, and and these central actors vary in type, i.e., civilian, military, etc \citep{Gattiker_2006}.  Additionally, structural similarities within a network of conflict dyads among countries have even been used to describe international conflict patterns \citep{Maoz_2006}.  In American politics, structurally descriptive methods have been used to identify the most influential members of the U.S. Congress based on their co-sponsorship networks \citep{Fowler_2006}.

In many applications of network analysis in political science authors attempt to explain a behavior or observed outcome from the structural features of the network being studied.  In the example of the co-sponsorship network study, the position of the legislators inferred by their centrality is used to predict the number of legislative amendments proposed by members.  This type of inference is common in descriptive network analysis.  It is difficult, however, to know the direction of causality.  Structurally descriptive work is often based on static snapshots of network data, or large aggregations over time.  As is the case in the aforementioned legislator study, which aggregates co-sponsorship data from 1973-2004.  From this perspective it is difficult to parse whether legislative influence comes as a result of co-sponsorship behavior, or that co-sponsorship behavior is affected by influence.  The same can be said for the other studies.  In each case, and in most examples of structurally descriptive work both in and outside of political science, the network is given. 

Descriptive techniques rarely give any insight as to the data generation process that resulted in the network being studied \footnote{It should be noted that in \citep{Gattiker_2006} the authors explicitly point out the limitations of structurally descriptive techniques for analyzing network time-series.}.  An alternative viewpoint is to assume that network structure is endogenous to the preferences of actors.  Then, estimate how various structural features or actor attributes contribute to network structure.  As a simple example, consider membership in a terrorist network.  One may assume that members of al-Qaeda were more likely to make ties with other known al-Qaeda affiliates.  Using a statistical technique known as exponential random graph models (ERGM) one could estimate a coefficient to measure this in- and out-group effect given the appropriate data \citep{Hunter_2008, Robins_2007, Snijders_2006}.  This type of network analysis treats networks as the dependent variable, and is the second common application of network analysis in political science.  

The innovation presented by ERGM models is disentangling the highly interrelated dependency structures of network relationships in order to generate unbiased estimates of network parameters.  The application of exponential random graph models, however, has only very recently emerged within political science.  Some work goes only so far as to suggest that ERGM models be used to correct for the dependencies present in international relations networks \citep{Hafner_Burton_2009}.  There is, however, some very recent research on causality on networks wherein the authors use an ERGM design to determine if there is systematic bias in reported contacts among lobbyists \citep{Fowler_2011}.  In fact, most of the application of ERGM models in political science remain unpublished working papers.\footnote{For further examples of working papers using this technique see the recent submission to the Political Networks section of APSA (\url{http://www.apsanet.org/content_69102.cfm}).}  Given the current research trends ERGM models will continue to proliferate the discipline.

A clear advantage of these techniques over structurally descriptive analyses is that the direction of the effect being measured is explicit.  Consider again the example of inferring the role of members of the U.S. Congress based on their structural position.  Using an ERGM design the problem is approached in reverse.  The model estimates separate coefficients for the centrality metrics and number of amendments from each members given the co-sponsorship network.  From this, one could potentially determine which had a larger or more significant effect on the structure of the network.  It is much more difficult, however, to model how these parameters will affect the nature of co-sponsorship relationships among legislators going forward.  

Analyzing longitudinal networks is a small but vibrant area of research within the network analysis community.  Much of this work has been done by Tom A.B. Snijders, and has focused on generating maximum likelihood estimators (MLE) for network panel data \citep{complexity_2009,Snijders_2010_2,Snijders_2010}.  These dynamic network models make two assumptions.  First, each actor is a singular agent determining its own relationships.  Second, these relationships are a function of a time-varying continuous random variable.   Then, a MLE is specified to estimate future states of the network given time interval network data, i.e., the structure of the network at $t_{0}, t_{1}, ...,$ and so on.  These models are extremely powerful, and have pushed the envelope of this discipline by incorporating network dynamics to a discipline dominated by static models.  Unfortunately, given the current state-of-the-art, these methods also have significant limitations, particularly with respect to variations in the type and availability of network data.

At present, there does not exists a flexible set of tools for modeling the growth and change of networks over time.  ERGM and dynamic MLE methods have introduced many new tools, however, methods for generating random networks from sparse and heterogeneous network data remain elusive.  Such techniques would be extremely valuable to social science research---and especially political science.

Suppose a researcher was interested in studying recruitment into an organization engaged in elicit activity, such as a terrorist group.  In view of the literature on recruitment into terrorist organizations \citep{Sageman_2008,Hegghammer_2006}, one could assume that individuals' social networks play a large role in the evolution of these groups.  As a matter of fact, gathering survey data from terrorists is difficult or impossible.  There does exists, however, sparse historical data on terrorist networks \citep{Carley_1998,Krebs_2002}.  Using this limited data, it may be useful to test theories of terrorist recruitment given the roles and positions of the actors in the network.  Alternatively, one might simulate a terrorist network and assign actor roles as some random or stochastic process.  Then, as with the real data, attempt to study recruitment within the networks.  Unfortunately, given the current set of tools this is not possible.

Ideally,  one could possibly learn something about the state of this organization from the natures of its structure and the type of actors in it.  Then, posit a model for change and simulate future states of the network.  In the case terrorist recruitment, a model of recruitment might assume that new members join as a function of specific personal attributes and always at the periphery. With the inherent internal security concerns of terrorists this seems reasonable.  One could also assume that recruitment only occurs with actors assigned that role in the network, and thus growth and change is localized to these areas of the network.  With a method that was able to leverage data in this way both of these models could be tested, and the resulting simulations could be analyzed and compared.

There also exists a burgeoning literature on the effects of social networks on political outcomes and collective actions.  This research suggests a strong relationship between the two \citep{McClurg_2003,Scholz_2006,Siegel_2009}.  As such, it may be the case that individuals are using social networks to overcome problems inherent in collective action.  Informational or efficiency gains may be made through the repeating of certain network structures in various contexts.  There are however, many theoretical considerations for how social networks affect collective action \citep{Siegel_2009}.  Up to this point, much of the applied research in this area has focused on experimental work and cooperative games \citep{Jackson_2008, Easley_2010}.  A classic example from this literature is having actors vote on a color, such as red or blue.  Players are incentivized to match the color chosen by everyone in the network but can only see the votes of their immediate neighbors.  In this case, the networks are treated as exogenous and the experiments attempt to study the types of networks that reach equilibrium.  One may, however, be interested in studying how variation in votes casts alters the trajectory of network structure.  Rather than studying only the collective act of voting for a color, by using the network's structure and color votes one could specify a model for how this network grows as a function of these two variables.

In the following sections I propose a technique for modeling network evolution in precisely this way.  Using information ascertained from a base set of network relationships, the evolutionary process is modeled using a set of graph motifs, or small constituent components of a network.  The proposed method attempts to fill the void in network modeling tools for social scientists elaborated above.  The remainder of this paper proceeds as follows.  First, a motivation for why using graph motifs in this way is a useful approach for modeling network growth.  Next, the formal specification of the graph motif modeling technique is described.  This abstract framework has been implemented in a software package.  This software is then used to provide a proof of concept for graph motif modeling.  This is done by recovering three classic random graph models: the Erd\H{o}s-R\`{e}nyi binomial random graph, the Watts-Strogatz ``small world'', and the Barab\'{a}si-Albert preferential attachment.  In the final section, the results of these simulations are discussed, noting strength and weakness of the technique.

\subsection*{Modeling random network growth} 
\label{sub:modeling_random_network_growth}

The research and development of random graph models that consistently characterize structural phenomena observed empirically in social and complex networks dates back to the work of Paul Erd\H{o}s and Alfr\`{e}d R\`{e}nyi in their seminal work on binomial random graph models \citep{erds_random_1959}.  In the intervening decades there has been an explosion of research in this area.  The so-called ``small-world'' network model was introduced by Watts and Strogatz \citep{watts_collective_1998}, and was predicated on two important observations in social networks: short average path length between nodes and a high level of localized clustering.  These structural phenomena were often observed in relatively small networks, but as technology improved so did the ability to study large complex networks.  Following the Watts-Strogatz model was the work of Barab\'{a}si and Albert \citep{albert_statistical_2002}, which noted that structure within complex networks exhibited ``preferential attachment,'' meaning a limited number of nodes drew in disproportionally more edges than the vast majority of others.  This process generated networks with ``heavy-tailed'' degree distributions \citep{barabasi_emergence_1999,albert_statistical_2002,Clauset_2009}.  Later, these models will be used as examples of the types of structures that can be modeled using graph motifs.

As mentioned, ERGM techniques have emerged as preferred method for modeling networks.  Random networks can also be generated with these techniques using Markov-Chain Monte Carlo (MCMC) simulations.  Rather than estimating coefficients for some network parameters given a network, random graphs can be generated given estimated parameters.  This class of models retains the structural consistency of previously developed models.  ERGM, however, assume a fixed number of nodes, and structure is modeled as random variables in a stochastic process \citep{Robins_2007}.

In addition to these general models, over the past several years an explosion of highly tailored network models have appeared.  These models were developed to address specific structural features of networks.  Some of these models are more closely related to graph theory, which attempt to bridge the gap between the classical concepts of Erd\H{o}s and Renyi with observed features in social networks \citep{Bollobas_2001,Newman_2003}.  Similarly, an alternative class of contemporary models takes an agent-level approach. These simulate structural growth as a decision process occurring endogenously through the nodes themselves \citep{Leskovec_2005,Snijders_2010}.  Literature on random graph models has provided enormous insight into the general structural dynamics of networks, they remain limited. Specifically, in both their underlying assumption about the means by which structure is generated and the types of networks these models can simulate.

Using a model specifically designed to approximate the dynamics of interest may be adequate.  The rigidity of these models, however, makes them much less useful for modeling poorly understood network dynamics.  The ERGM class of models can theoretically model any countable graph, which itself constitutes a monumental and unifying contribution.  In practice, however, the MCMC simulations used to generated random graphs from ERGM models achieve a much sparser set of graphs. The problem of model degeneracy is well known in the ERGM literature, and attempts have been made to address these problems \citep{dynamic_social_network}.  Unfortunately, the practical implications are quite limiting. With many models of interest degenerating into complete (fully connected) or empty graphs (no edges).

The primary shortcoming of these models is their treatment of the atomistic component of a network---the node.  In all of the models mentioned above, and in fact in the vast majority of random graph models of social networks, actors are modeled as entering the system in a vacuum.  That is, nodes enter free of any pre-existing structure.  In real networks, however, this is not the case.  Except in the simplest of cases, whenever an actor enters a network system that actor is bringing some degree of exogenous structure.  This structure will have an immediate impact on the growth trajectory of that network.  This is particularly true of human social networks, which exist in a rich, complex, and often hidden fabric of social ties.

Consider the network dynamics when two people meet each other for the first time.  Upon meeting, these individuals have changed the structure of their social networks by creating an edge between them. With that structure they have also brought with them their pre-existing social structure.  All of the people they already know: friends, family, co-workers, competitors, etc.  This meeting has not simply created a dyad existing in isolation, but rather it has connected two large components.  It may have also increased the probability that the single bridge created by this dyad will in turn become a cluster of shared relationships.  Figure 1 visually depicts the difference between these concepts.

\begin{figure}[ht]
    \begin{center}
        \subfigure[Dyadic model]{\includegraphics[width=6cm]{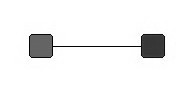}}
        \subfigure[Motif model]{\includegraphics[width=6cm]{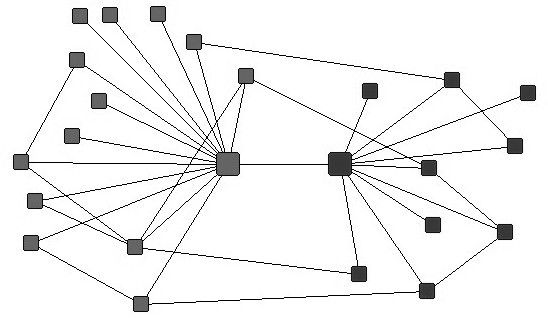}}
    \end{center}
    \caption{Competing models of social interaction}
    \label{fig:social}
\end{figure}

There is considerable nuance and ambiguity with respect to how to model the relationships in the right panel of Figure 1.  Instead, the dyadic relationship on the left is simply a binary event. The dependencies related to these ties can be a function of structurally descriptive metrics.  These could include network-level metrics, such as diameter, centralization, or density. Likewise, node- and edge-level attributes could be used, such as centrality metrics or node type.  The plethora of potential modeling parameters have led to a literature full of rigorous, yet limited models for network growth.  Current random graph models of social networks are useful, but are limited by oversimplified assumptions that ignore the inherent complexities of social structure.  

This research attempts to close the gap between the theoretical assumptions of current models and the self-evident reality of natural network interactions by providing a more flexible framework capable of modeling a much larger set of networks by leveraging graph motifs.

\section*{Graph motif model} 
\label{sec:natural_random_graph_model}

To overcome the limitations of current random graph models of social networks the concept of a graph motif model (GMM) is introduced.  This new framework is predicated on two key assumptions that distinguish it from other network modeling techniques.  First, new actors entering a network do not do so in isolation.  That is, actors bring exogenous structure to a network when entering it.  Models of social networks, therefore, should build new structure in an analogous way.  To model networks this way it is necessary to posit assumptions about these exogenous structures and the process by which they will enter the network.  For GMM the assumption is that new network structure will resemble currently observable structure in type and frequency.

With this assumption, current structure can be used to form the necessary beliefs.  This, however, forces a strict requirement for GMM that is not shared by other random graph models.  Specifically, the need for some base structure from which to derive beliefs about the network being modeled.  One could argue that all random graph models require base structure in that they all require some fixed number of nodes to model.  A set of nodes without structure still constitutes a base graph, despite its degenerate form.  This is particularly true of the Barab\'{a}si-Albert model of preferential attachment, which always begins with the same base structure.\footnote{In practice this is often modeled as a single dyad or a three-node line graph.}

The observation that networks perpetuate self-similarity as they grow has been noted several times in the empirical literature.  In fact, complex networks exhibit significant fractal scaling \citep{song_2005,kim_fractality_2006,kim_037102}.  That is, as their size increases, so too does the amount of self-similarity.  This observation forms the critical bridge between the first and second assumptions.  To be clear, there are several assumptions that could be used to form beliefs about network structure.  These include those mentioned earlier, such as node-level metrics, stochastic processes, etc.  Self-similarity is preferable because its empirically supported and not dependent on a graph's type.  If a model were based on a node-level metric it could only describe networks for which that metric was relevant.  For example, many metrics are only defined for directed or undirected graphs, weighted or unweighted graphs, weakly connected or strongly connected graphs, and so on.

The GMM framework described here applies to undirected and directed graphs with an arbitrary set of node or edge attributes.  While this allows for an extremely rich set of possible models, it precludes some graph forms.  Specifically, this restricts the proposed model from describing multigraphs and hypergraphs.  It may be possible to incorporate multigraph models into the GMM framework; however, the abstract nature of hypergraphs makes their applicability unclear.  For example, consider a hypergraph wherein a single edge is incident on many nodes.  This is not a construction that models social interactions naturally, and thus incorporating them into the model has limited value.\footnote{A ``multigraph'' is defined as a graph where any two nodes may have multiple edges between them.  Conversely, a ``hypergraph'' is defined as graph where a single edge may be connected on any number of nodes.}  

With these restrictions, the model proceeds as follows: require some graph $G$ of arbitrary size and some integer $\tau>1$. Next, count all of the subgraph isomorphisms in $G$ of graphs $i\in I$, where $I$ is the set of all \emph{single-component graphs formed by $\tau$ nodes}.  The restriction that $\tau$ be strictly greater than one accounts for the fact that a $\tau\ge1$ would allow for a singleton element.  Allowing this would violate the first assumption of GMM for exogenous network structure.  These single component graphs are the motifs on which the entire GMM framework rests.   For example, suppose $\tau=3$, then $I=[\{V=2,E=1\}, \{V=3,E=2\},\{V=3,E=3\}]$ where $V$ is the number of nodes and $E$ the number of edges for graph $i\in I$.  In this example $\{V=3,E=1\}\notin I$, as this graph contains two components: a dyad and an isolate.  Also, note that these motifs have a natural ordering given their number of nodes and edges.  This ordering will become critical to how new network structure enters the model.

Next, let $f(i_n,G)$ be a function that describes the number of subgraph isomorphisms of $i_n$ contained in $G$.  Then, let $S$ be an ordered $n$-tuple where $S=\{i_1,i_2,...,i_n\}$, such that $i_n$ is increasing in number of nodes and edges.  For two graphs to be isomorphic there must be a one-to-one correspondence among the nodes and edges of two graphs.  A subgraph isomorphism between two graphs \emph{G} and \emph{H}, therefore, is defined as such a correspondence for graph \emph{G} in an induced subgraph of \emph{H}.  This construction is very useful, as it allows for the quantification of motif frequency in any given base structure.  Put another way, this is the composition of a graph given some set of possible constituent parts. While the subgraph isomorphism problem is known to be NP-complete, certain cases can be solved in polynomial time and several algorithmic approximations have been proposed \citep{ullmann_algorithm_1976}.\footnote{The term ``NP-complete'' refers to the complexity class of a problem.  Specifically, an NP-complete problem is one for which a solution can be verified in nondeterministic polynomial time.  Simply, this represents a very difficult problem to solve algorithmically.}  The ordered tuple $S$ can then be used to generate beliefs about the type of structure entering the network as it grows in size and complexity.  In order to generate these beliefs some function must be defined over the structures in $S$.

\subsection*{Generating beliefs about structure} 
\label{sub:generating_beliefs_about_structure}


A critical aspect of modeling network growth using motifs is specifying what types of motifs are evolving the graph. Subgraph isomorphism provides a means for calculating the frequency of these motifs.  It is still necessary, however, to specify how these discrete counts are used to generate the beliefs about the network's evolution.  A straightforward way to do so is to simply define a discrete probability mass function over these counts.  

Given $S$, define a probability mass function (PMF) such that $\sum_i^{S}Pr(X=i)=1$, where $i$ is element of the tuple $S$ with discrete probability.  The sum of probabilities for all elements in $S$ is equal to one.  As the number of elements in $S$ is dependent on $\tau$, it is not necessary that the PMF relate exclusively to the elements of $S$.  For the purposes of models specified in this paper the PMF defined are both exclusive to this set, i.e., only account for motifs defined by $\tau$.  For example, recall the simple set of motifs described for $\tau=3$.  In this case, a GMM would not require a PMF that satisfied the above requirement for the complete graph mode of four nodes because this motif would be excluded from $S$ by definition.  

This allows for a large set of possible PMF to determine the probability a given motif will enter the network.  This function may rely explicitly on the subgraph isomorphism counts, wherein zero probability mass is defined for any motif with no subgraph isomorphism in the base structure.  Alternatively, it is also possible to specify a PMF that models the probability of motifs as a discrete probability distribution over all elements of $S$.  Below I describe two examples of PMF that could be used in a GMM.  The first is an explicit function over the elements in $S$, while the second function provides positive probability mass for all elements of $S$.  In the latter PMF, regardless of whether a given motif was observed as subgraph isomorphisms in the base structure it may still have positive probability mass.

\begin{figure}[ht]
    \begin{equation}
        F(i)=\frac{S_i}{\displaystyle\sum_{n=1}^{S}S_n}
    \end{equation}
    \caption{Explicit PMF for motif probability}
\end{figure}

Equation 1 provides a discrete probability mass over $S$.  This function states that the probability $i_n$ will be the next structural component of $G$ is the proportion of subgraph isomorphisms found for $i_n\in G$, normalized by the total number of subgraph isomorphisms counted $\forall i\in S$. $F$ thus provides the necessary prior beliefs to generate new structure in $G$.  Again, this function will assign zero probability mass to any motifs that are not observed as subgraph isomorphisms in the base structure.  This can be problematic, as it presumes that certain motifs will never enter a graph.  This also clearly limits the possible networks it can model using this PMF.  In other cases, however, this limitation may be necessary.  For example, in bipartite networks certain motif structures cannot exist in order to maintain the bipartite structure.\footnote{Bipartite networks are defined as have two mutually exclusive node sets, wherein edges can only be formed between nodes from different sets.}

It may also be useful, therefore, to have a PMF that assigns positive probability to all motifs.  Here, I utilize the natural ordering of elements in $S$ by their structural complexity to fit a canonical discrete probability distribution to the set of motifs.  Specifically, I define an alternative PMF for the elements of $S$ in terms of the Poisson distribution in Equation 2 below.

\begin{figure}[ht]
    \begin{equation}
        F(i;\lambda)=\frac{\lambda^{i}e^{-\lambda}}{i!}
    \end{equation}
    \caption{Poisson PMF for motif probability}
\end{figure}

In this specification the ``mean'' of the distribution, represented by the shape parameter $\lambda$, is the mean of all motif counts in $S$.  The natural ordering of motifs by complexity fit the motivation of the Poisson distribution to model event counts.  Here I consider the occurrence of increasingly complex motifs within a given base structure as an increasingly rare event.  Likewise, the most likely motifs to enter a graph may have probabilities centered around the motif with mean complexity in the base structure.  If these assumptions do not reflect the data generating process present in the base structure, however, such a specification is misplaced and an alternative specification of the PMF should be used.

It is important to note that these, or any PMF defined over $S$, have a direct effect on the nature of the GMM specified.  The function defined in Equation 1 requires the least amount of assumption about the probability of motifs in the model.  It relies explicitly on subgraph isomorphism counts.  This, however, is limiting and alternative methods may be desirable.  As such, it may be useful to define a PMF from the canon of discrete probability mass functions.  I have done this using the Poisson distribution in Equation 2.  In each case, the given assumption must be relevant to the network growth process being modeled.  Once probabilities are defined over $S$, the next step in specifying a GMM is to define methods for adding these structures to the graphs.  This function must map motifs from $S$ into the base structure.  As before, this function can take many forms.

\subsection*{Generating new structure} 
\label{sub:rules_for_growth_and_termination}


The next step in the model is to draw some motif from $S$ using the probability distribution and add it to the network structure by some growth rule.  This rule is denoted $R(\cdot)$ and is defined as a mapping $R:i_n\rightarrow G$, which is restricted only by the graph theoretic constructs assumed by $G$.  Specifically, the decision rule must be applicable to the fundamental constructs of $G$ and subgraph elements of $S$, but is otherwise open to the particularities of a model's design.  For example, a growth rule cannot assume a multigraph as input because this graph type is restricted from the GMM framework.  In the following section a simple growth rule is defined.

After each iteration of growth the process for forming structural beliefs is repeated, and the probability distribution recalculated.  The continual updating of beliefs as the network grows allows for a certain degree of path dependance in the model, as probability mass may converge to the most likely motifs as the network grows.  This may or may not be viewed as an advantage of the model, but future versions will allow for both static and dynamic probability calculations over $S$.  This process continues until the model has satisfied some termination rule. This is denoted as $T(\cdot)$, and is restricted as $R(\cdot)$.  

The means by which the evolutionary process is modeled are intentionally left open.  The GMM framework is meant to support any number of possible growth models.  Beyond the few restrictions described above, the choice and growth and termination functions are completely at the discretion of the modeler.  This is a dramatic philosophical departure from both the closed form models and stochastic models discussed earlier.  Unlike these strict models the GMM framework retains flexibility.  This allows the technique to describe a large and nuanced set of graphs.  In the following section the algorithmic implementation of this method is described in detail, and one simple implementation of a GMM is specified.  Before proceeding, however, here is a review of the core elements of the method for modeling networks using graph motifs.

The framework for modeling network structure using graph motifs described above attempts to overcome the limitations of current methods by proposing a flexible modeling framework wherein a rich set of graphs can be described.  This is motivated by the need to not only describe the static structure of networks, but also model how they evolve over time.  To achieve this, two key assumption are made.  First, some base structure upon which to form beliefs about the type of graph being modeled must be provided.  Then, the constituent parts of this base structure---graph motifs---represent a useful proxy for the data generating process present in the network being modeled.  Again, these assumptions are in stark contrast to those of many traditional network models.

\begin{figure}[H]
    \begin{enumerate}
        \item Require some base graph $G$ of arbitrary complexity
        \item Given some integer $\tau>1$, the set $I$ contains all single-component subgraphs formed by $\tau$ nodes
        \item Define $S$ as an ordered $n$-tuple containing all $i\in I$
        \item Define the function $f(i_n)$ to count the number of subgraph isomorphisms of $i_n\in G$, and a PMF over all elements in $S$
        \item Draw structure from this probability distribution and add that structure to the network by some growth rule $R(\cdot)$
        \item Repeat steps 4-5 until the some termination rule $T(\cdot)$ is satisfied 
    \end{enumerate}
    \caption{The basic steps of the GMM framework}
\end{figure}

The GMM framework brings with it a different set of limitations, many of which will be discussed in the conclusion.  As much of the model hinges on subgraph isomorphisms counts, this model requires a sophisticated computational implementation.  In the following section an implementation is described using the \texttt{Python} programming language to develop the \texttt{GMM} package for graph motif modeling.

\section*{Algorithmic implementation: the \texttt{GMM} \texttt{Python} Package} 
\label{sec:implementing_the_gmm}

Before any implementation of the GMM can proceed, it will be necessary to have a means for representing complex networks computationally in the \texttt{Python} language.\footnote{For more information on the Python language see \url{http://www.python.org/}}  Fortunately, the \texttt{NetworkX} package is a highly-developed Python package for the creation, manipulation, and study of the structure, dynamics, and functions of complex networks \citep{Hagberg_2008}.\footnote{The \texttt{NetworkX} package exploits existing code from high-quality legacy software in \texttt{C}, \texttt{C++}, \texttt{Fortran}, etc., is open-source, and fully unit-tested.  For more information on NetworkX see \url{http://networkx.lanl.gov/}}  \texttt{NetworkX} is capable of representing graphs of arbitrary complexity, including both node and edge attribute data.  These features make \texttt{NetworkX} ideally suited as the computational foundation for an algorithmic implementation of the GMM framework.

The \texttt{GMM} package consists of two object classes.  The first is the ``gmm'' class itself, which is the essential element of any model. This  requires three arguments: a \texttt{NetworkX} graph object as the model's base structure, and special \texttt{Python} functions as growth and termination rules.  With these elements in place, the ``gmm'' object can be used to simulate the evolution of the given base structure.  The remaining scaffolding built into this class is in place to verify that all model parameters are valid to a GMM model, store these parameters appropriately, and provide functionality for storing and retrieving information about a given GMM simulation.

\begin{figure}[ht]
    \centering
    \includegraphics[width=10cm]{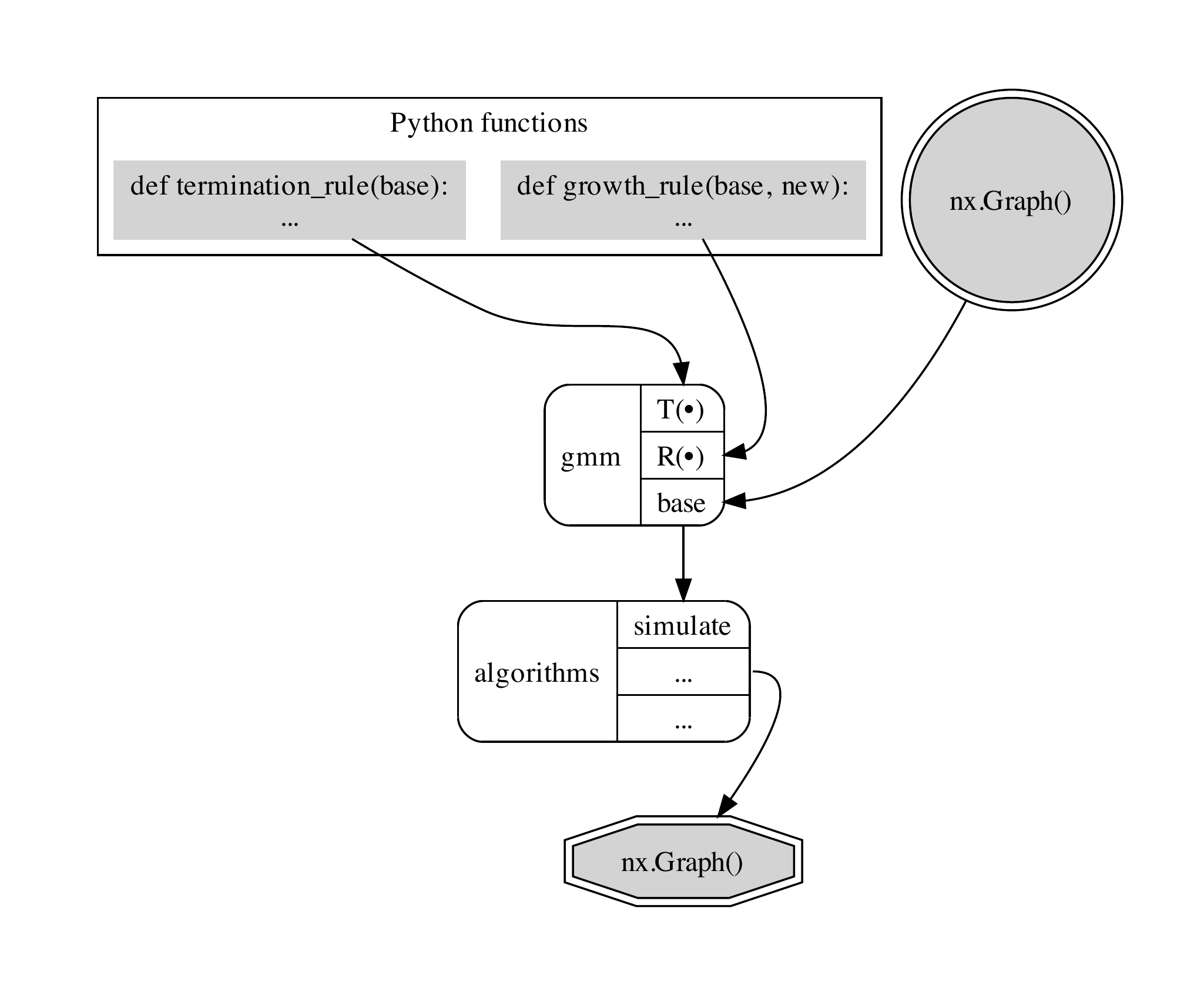}
    \caption{Implementation of \texttt{GMM} with dependencies}
\end{figure}

The second class is ``algorithms,'' which provide all of the functionality for properly running a simulation and generating network structure from a given ``gmm'' object.  This contains all of the functions needed to create a set of graph motifs, generate beliefs about how that structure enters the model, and the actual generation of new network structure.  With respect to generating beliefs, the two PMF discussed in the previous section are included by default.  The Poisson function requires \texttt{SciPy}---a third-party \texttt{Python} scientific computing package---to generate the correct probabilities. \texttt{NetworkX} also requires this package as a dependency; therefore, its use in the \texttt{GMM} package does not compound software requirements.\footnote{For more information on \texttt{SciPy} see \url{http://www.scipy.org/}} 

Finally, as stated previously, the subgraph isomorphism problem is known to be NP-complete and requires a sophisticated approximation.  In this case, the VF2 algorithm---the most commonly used algorithm to evaluate subgraph isomorphism---is included in \texttt{NetworkX} and is used to perform the necessary calculation for matching subgraph isomorphism \citep{Junttila_2007,Cordella_2001}.  With these two simple classes, it is possible to specify a rich set of GMM.  Figure 6 above illustrates the basic computational framework described here.

Following the example of high quality scientific \texttt{Python} packages, such as \texttt{NetworkX} and \texttt{SciPy}, the \texttt{GMM} package is open-source and fully unit-tested.  All of the code is free to inspect and download at this website: \url{https://github.com/drewconway/gmm}, which includes all unit-tests to verify all function's execution.  The software is also listed in \texttt{Python's} official package index at \url{http://pypi.python.org/pypi/GMM}.  Detailed descriptions of the algorithms, their requirements, and additional examples are provided in the \texttt{GMM} package documentation, which can be viewed here: \url{http://www.drewconway.com/gmm/}.  In the next section a very basic GMM model is specified and simulated using this software.

\subsection*{A simple GMM with random growth} 
\label{sub:a_simple_gmm_random_growth}

As described, the first step in specifying a GMM are to determine the base structure.  Then, the growth and termination rules to be used to simulate growth.  In this example I use the canonical Petersen graph as the base structure, and two very simple rules.\footnote{The Petersen graph, often denoted as $K_{10}$, is a ten node graph with uniform degree of three.  It is a well-studied graph for its many known properties, such as being a non-planar.}  The termination rule will be a ``node ceiling,'' whereby the model will terminate growth once the network contains at least 250 nodes.  For growth a random attachment rule will is implemented. When a graph motif enters the graph, a random node from the motif will be connected to a random node from the current base structure.  These algorithms are implemented in pseudo-code below.

\begin{algorithm}[H]
    \begin{algorithmic}
        \REQUIRE $G$
        \IF {$G >= 250$}
            \RETURN \TRUE
        \ELSE
            \RETURN \FALSE
        \ENDIF
    \end{algorithmic}
    \caption{Pseudo-code ``node ceiling'' termination rule}
\end{algorithm}

\begin{algorithm}[H]
    \begin{algorithmic}
        \REQUIRE $G,H$
        \STATE $G=G[H]$
        \COMMENT{Compose $H$ with $G$} 
        \STATE $r_{1}=RANDOM(G)$; $r_{2}=RANDOM(H)$
        \COMMENT{Select random nodes from each graph}
        \STATE $G=EDGE(G,r_{1},r_{2})$
        \COMMENT{Create edge}
        \RETURN $G$
    \end{algorithmic}
    \caption{Pseudo-code random growth rule}
\end{algorithm}

This example is not a model of any particular network growth mechanism.  It is a useful example in that it shows the power of the GMM framework even with simple rules.  The Petersen graph is made of all closed motifs, i.e., there are no pendants or pendant chains present in the graph.  As such, the random growth rule connects these structures by single edges.  This mechanism will create simple chains of whatever motifs are drawn from the probability distributions.  Figure 6 below illustrates this, with the Petersen graph shown at the left, and the resulting simulation on the right.  The elongated structure in the right panel shows the chains of motifs growing in different directions, as the random selection of nodes caused growth to occur along several paths.

\begin{figure}[H]
    \centering
    \subfigure[Petersen graphs as base structure]{\includegraphics[width=7cm,clip,trim=2cm 6cm 4.5cm 6cm]{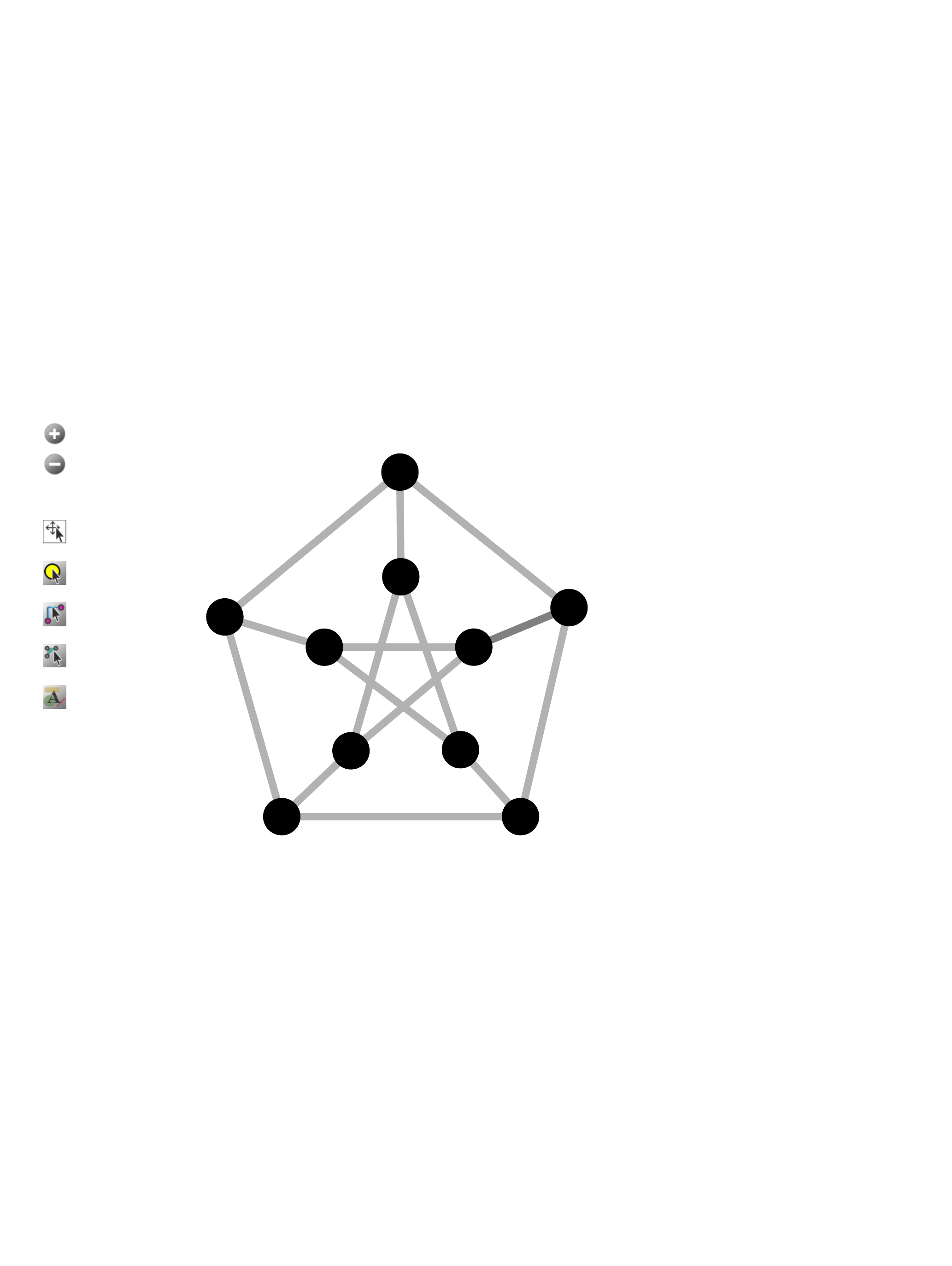}}
    \subfigure[Simulated GMM results]{\includegraphics[width=7cm,clip,trim=0cm 3cm 0cm 3cm]{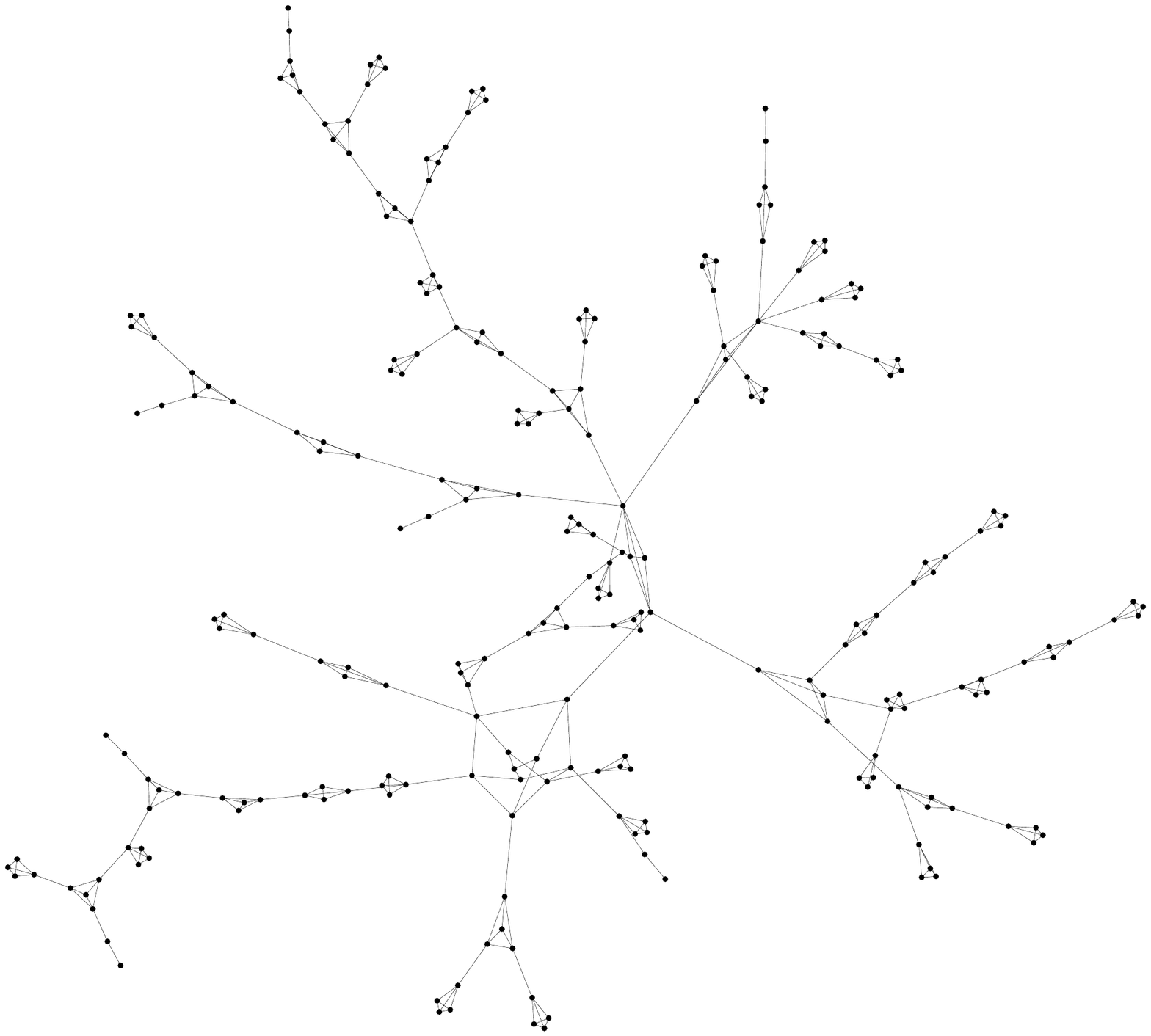}}
    \caption{Result of simple GMM with random growth rule}
\end{figure}

Using this same framework it is possible to model much more complex networks.  In the following section, three classic random graph models are recovered using the graph motif modeling method: the Erd\H{o}s-R\`{e}nyi binomial random graph, the Watts-Strogatz ``small world'' model, and the Barab\'{a}si-Albert preferential attachment model.  This exercise is meant as a proof of concept for GMM, which highlights many strengths and weaknesses of the technique.



\section*{Recovering classic random graph models with GMM} 
\label{sec:recovering_classic_random_graph_models_with_gmm}

To show the applicability of the GMM framework for modeling the growth of networks over time I propose three specifications that attempt to recover classic random graph models.  Each of these models has already been mentioned in earlier sections, and all are excellent benchmarks for GMM.  A primary reason for their utility is that each of the classic models described in this section attempts to model distinctly different network growth mechanisms.  As the experiments will illustrate, one of the most powerful features of the GMM technique is the ability to describe many different growth mechanism within a single modeling framework.

The first model these experiments attempt to recover is the most classic of all random graph models: the Erd\H{o}s-R\`{e}nyi (ER) binomial random graph model \citep{erds_random_1959}.  The model specification is very simple.  Given some number of nodes $n$, and a probability $p$ that any two nodes from that set form an edge.  This stochastic processes generates the random graph.  If $p=0.5$, then for each node a coin toss would determine if that node formed a connection to any of the other nodes.  The model is referred to as a binomial random graph because it produces networks with degree distributions that fit binomial distributions for the given $p$ and $n$.

To specify a GMM that will recover this model an appropriate base structure and growth rule will be needed.  The termination rule in this case will be a simple ``node ceiling'' because the ER models networks using a fixed number of nodes.  In fact, all of the classic random graph models discussed in this section use a fixed number of nodes.  As such, a node ceiling termination rule is used in all GMM described here.  The growth rule for a GMM equivalent of the ER model follows closely to the random growth rule specified in Algorithm 2.  The difference is that in this case there is an additional parameter $p$, which is the probability that the motif entering the network will form a tie to all of the nodes in the base graph.  For each node in the motif and each  node in the base structure, draw a random value from a uniform distribution on the unit interval.  If that value is less than or equal to $p$, create an edge between the node in the motif and the node in the base structure.  As before, a pseudo-code implementation of this algorithm is provided in Algortihm 3 below.

\begin{algorithm}[H]
    \begin{algorithmic}
        \REQUIRE $G,H,p$
        \STATE $G=G[H]$
        \COMMENT{Compose $H$ with $G$} 
        \FOR{$i$ in $H$}
            \FOR{$j$ in $G$}
                \STATE $ran=RANDOM(low=0, high=1)$
                \COMMENT{Draw a random value from a uniform distribution}
                \IF {$ran <= p$}
                    \STATE $G=EDGE(G,i,j)$
                    \COMMENT{If random draw less than $p$, create edge}
                \ENDIF
                \ENDFOR
                \ENDFOR
                \COMMENT{For each node in $H$ test if it will connect to each node in $G$}
        \RETURN $G$
    \end{algorithmic}
    \caption{Pseudo-code ER growth rule}
\end{algorithm}

\begin{figure}[H]
    \centering
    \subfigure[E-R ($p=0.5$, $n=50$)]{\label{fig:ER50}\includegraphics[width=.37\textwidth]{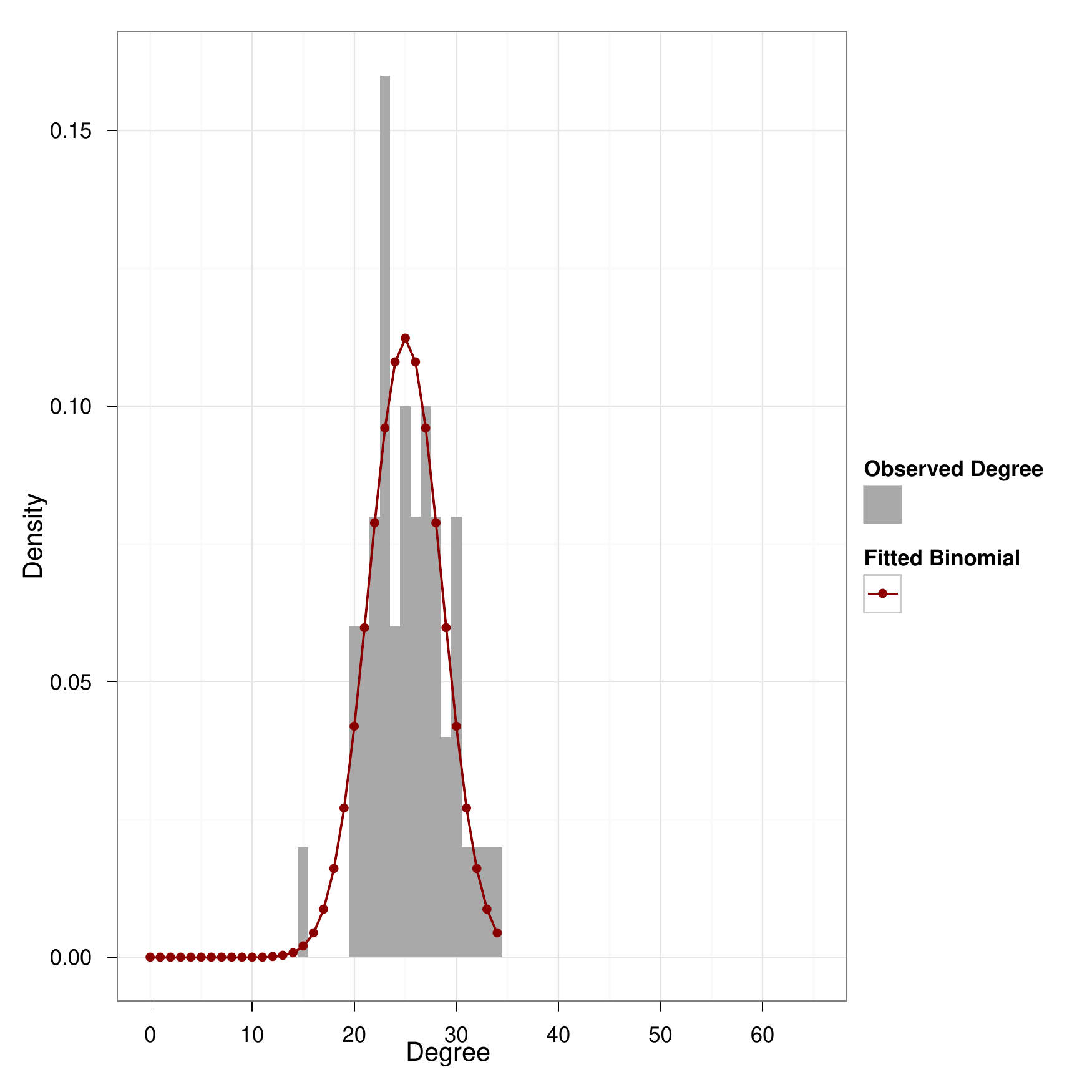}}
    \subfigure[GMM ($p=0.5$, $n=50$)]{\label{fig:GMM50}\includegraphics[width=.37\textwidth]{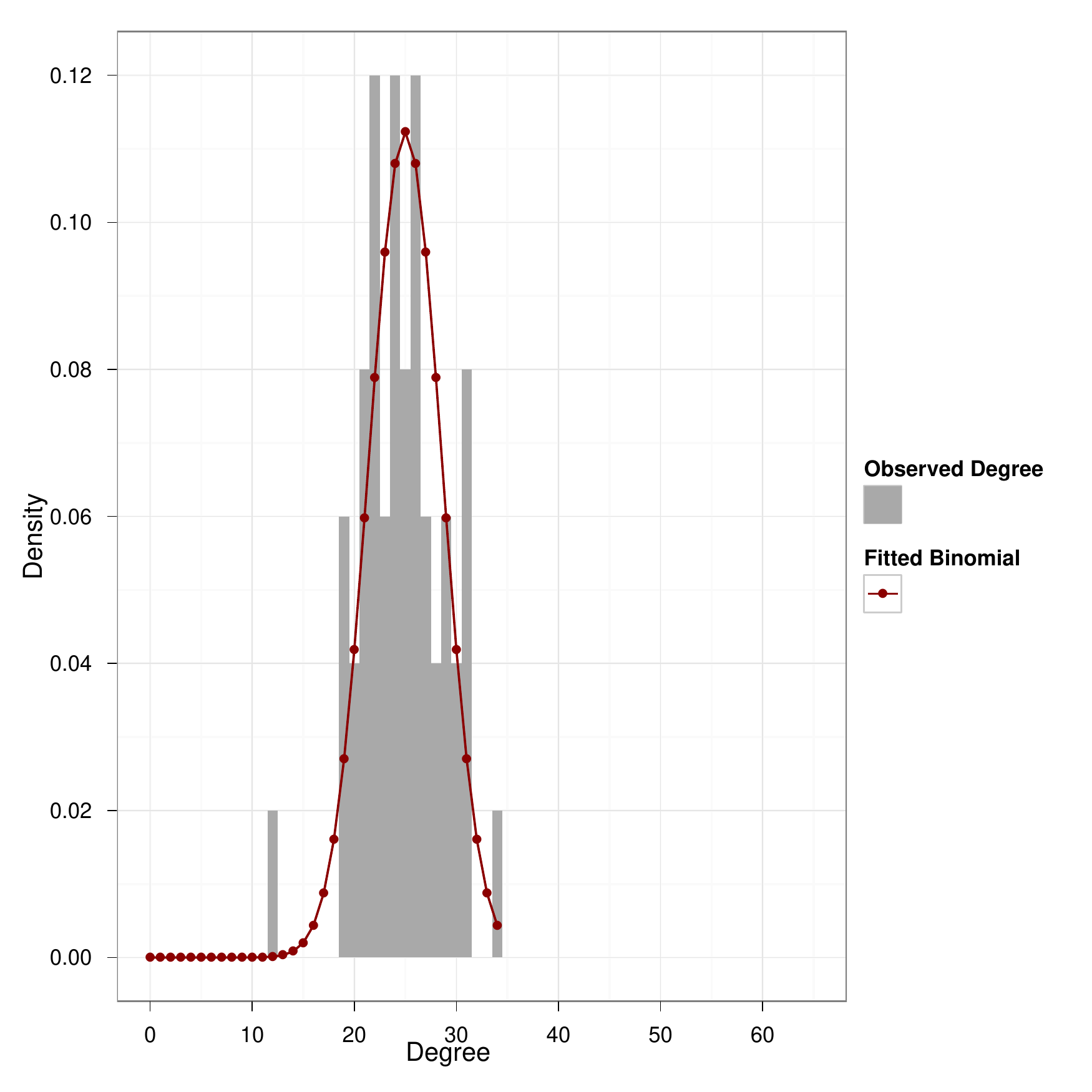}}
    \subfigure[E-R ($p=0.5$, $n=75$)]{\label{fig:ER75}\includegraphics[width=.37\textwidth]{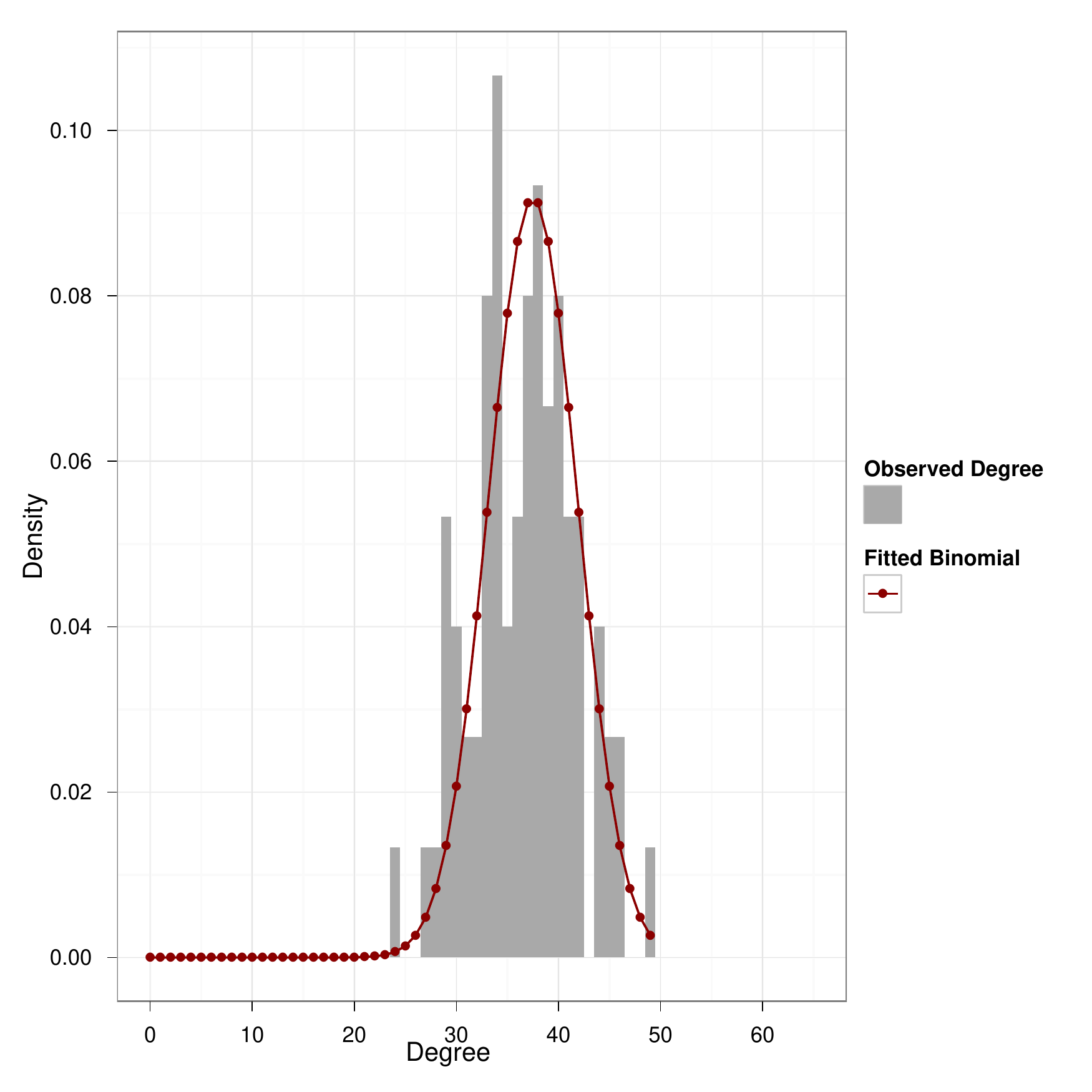}}
    \subfigure[GMM ($p=0.5$, $n=75$)]{\label{fig:GMM75}\includegraphics[width=.37\textwidth]{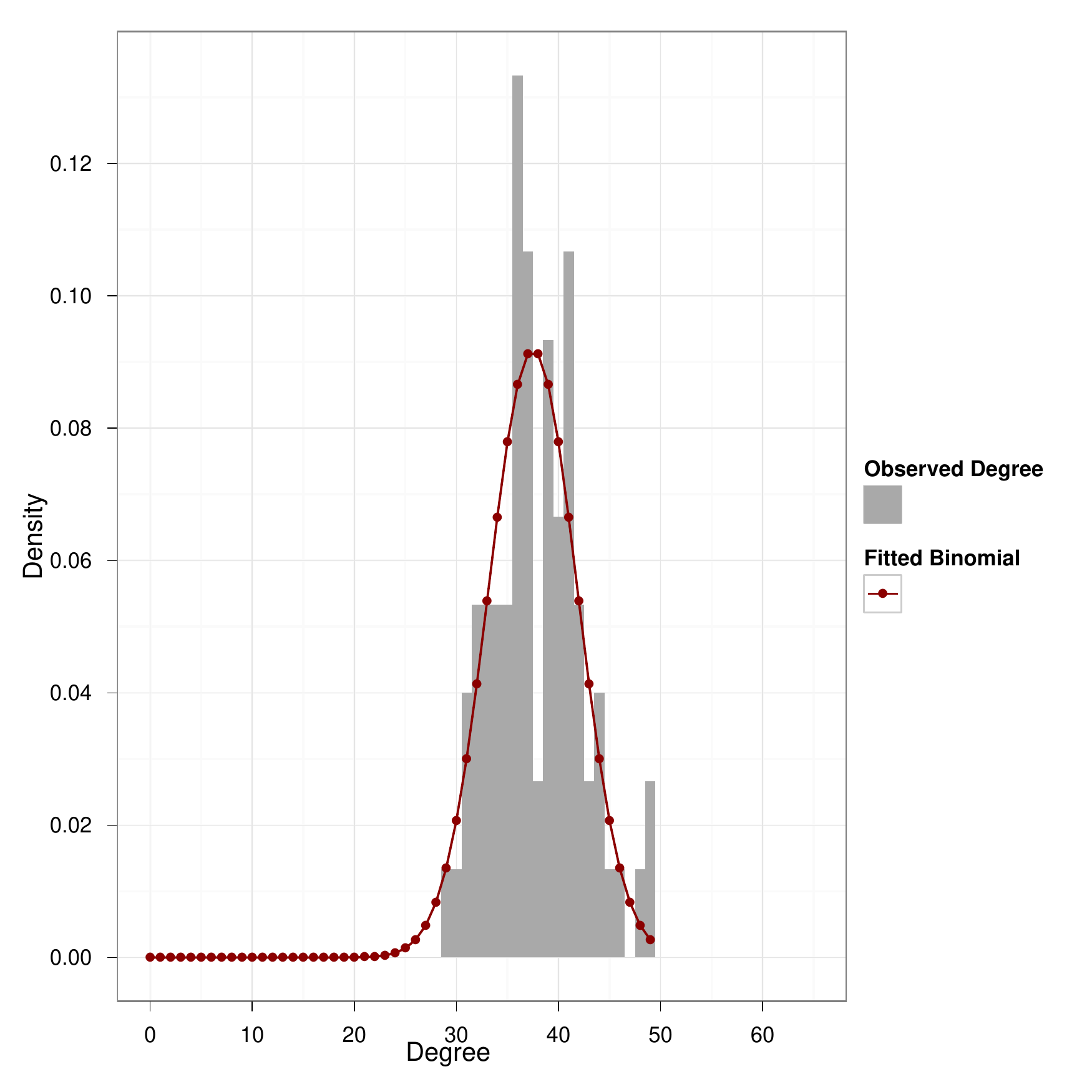}}
    \subfigure[E-R ($p=0.5$, $n=100$)]{\label{fig:ER100}\includegraphics[width=.37\textwidth]{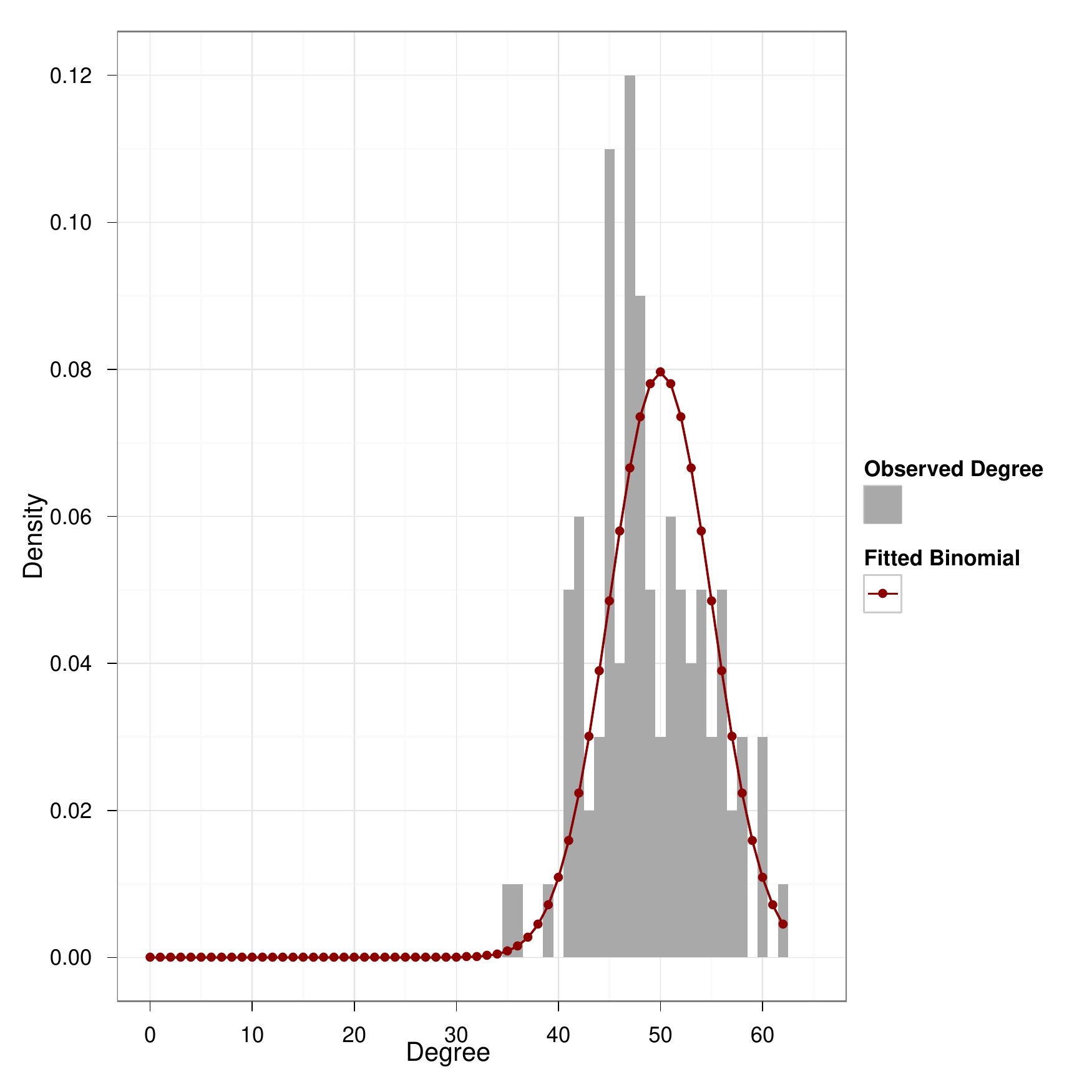}}
    \subfigure[GMM ($p=0.5$, $n=100$)]{\label{fig:GMM100}\includegraphics[width=.37\textwidth]{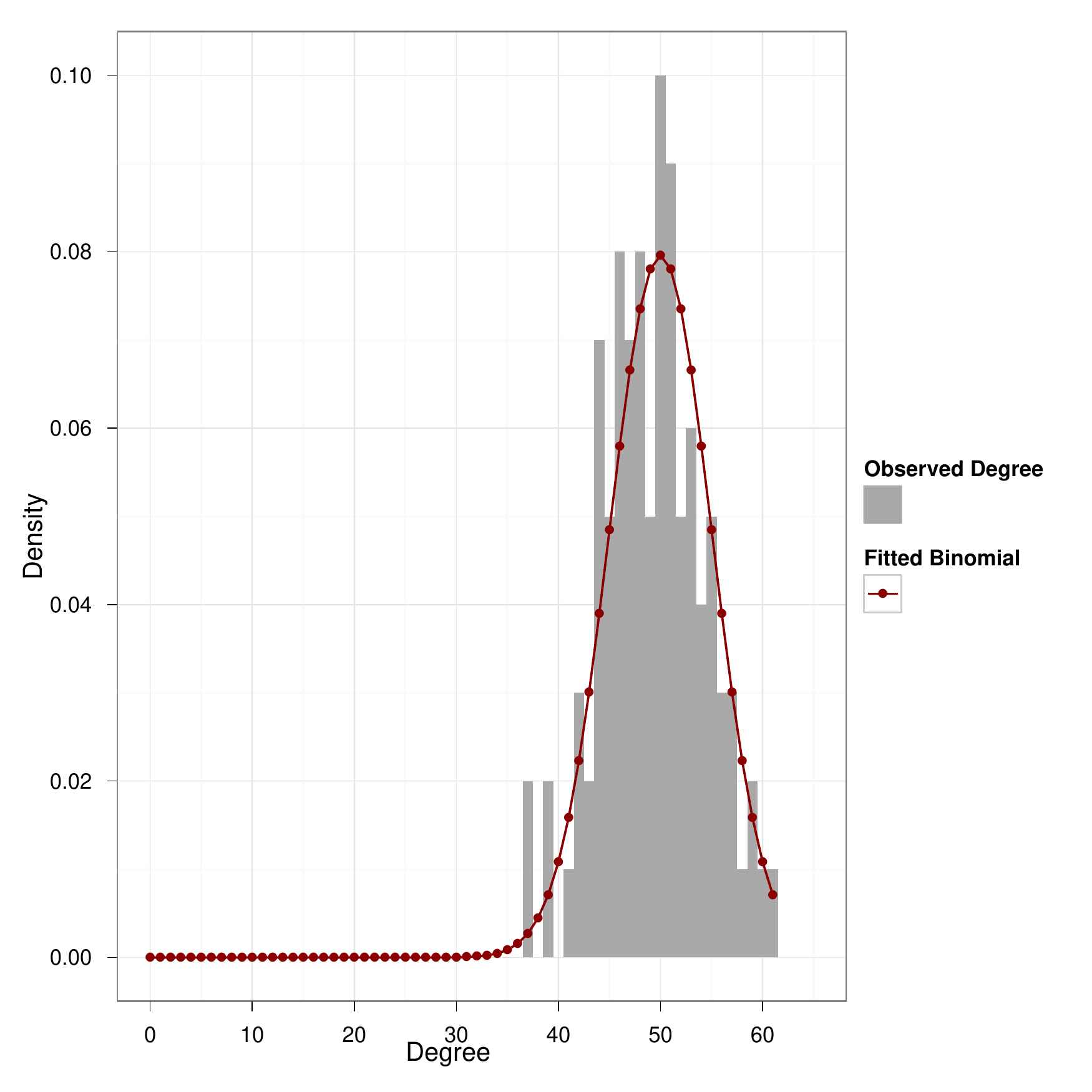}}
    \caption{Degree distribution and binomial fit for both classic Erd\H{o}s-R\`{e}nyi (ER) models and GMM equivalent.  The graphs in the left column represent degree distributions for classic models, all with $p=0.5$ and varying sizes with $n=\{50,75,100\}$.  In the right column are the degree distributions for simulated GMM graphs using a growth rule that attempts to mimic the E-R model.  The grey bars are the densities of observed degree in each network, and the connect red points are the fitted binomial densities.  In each case, $p=0.5$ and the base structure is a randomly generated ER graph with $n_{base}=n-25$.  For example, in panel (b) the base structure for the GMM was a random E-R with p=$0.5$ and $n=25$.  This pattern is consistent through all GMM simulations.}
    \label{fig:er-comp}
\end{figure}

To run this experiment I generated four classic ER random binomial networks.  These networks were generated using the built-in functionality of \texttt{NetworkX}. All subsequent classic models are also generated using this software.  For all networks $p=0.5$, but the number of nodes in each network increases from 25 to 100 at intervals of 25.  These networks are used as the benchmarks to evaluate the GMM simulations that follow.

To evaluate the result from the GMM simulations a measure of their ability to recover the data generating process described by the ER model will be needed.  The ER model describes a purely random data generating process.  With $p=0.5$, each node has an equal probability of creating an edge to every other node.  The degree distribution of the resulting networks, therefore, should follow a binomial distribution where the number of trials is the number of nodes minus one (no self-loops) and a probability of $0.5$.  The evaluation of the simulation results, and subsequent comparison to classic ER models, is based on the goodness of fit of the observed degree distributions to theoretical binomial distributions with the same specification.

Figure 7 illustrates these fits graphically.  In the left column of this figure are the degree distributions for all of the classic ER models, with network size increasing by 25 from top to bottom.  The grey bars in the figure are the observed degree distributions, and the connected red points are the theoretical binomial densities.  The GMM simulation for an equivalent network is the second column, with graph sizes matched left to right. As you can see from the these visualizations, both the classic ER model and the GMM equivalent produce degree distributions that well approximate the theoretical binomial densities.

To statistically compare the goodness of fit between the ER and GMM simulations a series of simple linear regressions are calculated wherein fitted binomial densities are regressed on observed degree distribution densities for all graphs.  A standard approach to testing the goodness of fit for count data is to use the Pearsons Chi-squared test.  In this case, however, due to an abundance of zeroes in the tails of the observed degree distributions this technique is not robust.  The linear models can be used to assess both the quality of the fit via the $R^{2}$ and root mean squared error (RMSE) values, as well as the quality of the models themselves using AIC.  The results are reported in the table in Figure 8.

\begin{figure}[H]
    \centering
    \begin{tabular}{lccccc}
        Graph & Binom. Density & Std. Error & $R^{2}$  & RMSE & AIC \\ \hline \hline
        ER ($n=50$) & 0.9339 & 0.0827 & 0.7880 & 0.0183 & -174.6894  \\
        GMM ($n=50$) & 0.9028 & 0.0819 & 0.7798 & 0.0181 & -175.3879 \\ \hline
        ER ($n=75$) &  0.8712 & 0.0638 & 0.7911 & 0.0133 & -283.8167 \\
        GMM ($n=75$) & 0.9717 & 0.0713 & 0.7901 & 0.0149 & -272.6114 \\ \hline
        ER ($n=100$) & 0.8943 & 0.0766 & 0.6859 & 0.0152 & -342.4095 \\
        GMM ($n=100$) & 0.9939 & 0.0416 & 0.9032 & 0.0082 & -412.7616 
    \end{tabular}
    \caption{This table reports the results of several linear models wherein fitted binomial densities are regressed on observed degree distribution densities.  This is done to measure the quality of fit for the observed degree distribution to a theoretical binomial for all graphs.  The coefficients for the theoretical densities are reported, with standard error.  The $R^{2}$ and root mean squared error (RMSE) values are reported as measures of the quality of fit, and AIC as a measure of model quality.  The p-values for all models are significant at the 0.99 level.}
\end{figure}

The table in Figure 8 reinforces the observation from Figure 7 that all simulations produce good fits.  In addition, these values provide some numerical basis for comparing the results of the ER and GMM simulations.  The most important observation from this table and the preceding graphs is that the GMM is able to successfully recover the classic ER model.  In all cases, the quality of fit for observed degree distribution is at least as good in the GMM simulations as the ER.  Only in the experiments where $n=100$ is the GMM model noticeably better.  This difference, however, would likely be diminished if cumulative comparisons were made across many simulations with the same parameterization.

These results are encouraging, as the GMM was able to recover very easily the results from the ER model.  That said, the ER does not describe a network evolutionary process that is observed naturally.  People do not form relationships at random.  A better test of the GMM is to attempt to recover a classic model that is designed to describe such a natural process.  The Watts-Strogatz  (WS) ``small-world'' model attempts to do this, and is the focus of the second set of experiments.

As mentioned in the introduction, the WS model is motivated by the observation the social networks often exhibit two structural features: short average path length between nodes and a high level of localized clustering \citep{watts_collective_1998}.  Short average path lengths indicate a network that is relatively dense, where any two nodes have few intervening connections between them.  High localized clustering is observed in networks with many cliques, where close nodes are densely connected.  Both of these features lead to social interactions that are associated with a small world effect, hence the name.

To produce a network of size $n$ and these structural features the WS model assumes a regular lattice of size $n$ and mean degree $k$.  This regular lattice is---in effect---the base structure used in the WS model.  In order to produce the small world effect an additional parameter $p$ is incorporated.  This is the probability that each edge in the regular lattice will be ``re-wired'' to some other random node in the network.  That is, given a regular $n,k$-lattice, the model iterates over each edge and rewires it with probability $p$.

This classic model achieves the desired result.  To measure localized clustering WS uses the mean clustering coeffecient for each graph.  Average shortest path length is used as the second measure, and both metrics are normalized by the equivalent measure for a regular $n,k$-lattice with no re-wiring.  This latter transformation is used to make the results comparable across variation in $p$ for fixed $n$ and $k$.  An interesting footnote of the original WS paper is that results were only reported for simulations on network with 10,000 nodes.  This is a relatively large number of nodes, given the theoretical motivation is to model smaller social groups.

\begin{figure}[H]
    \centering
    \includegraphics[width=.8\textwidth]{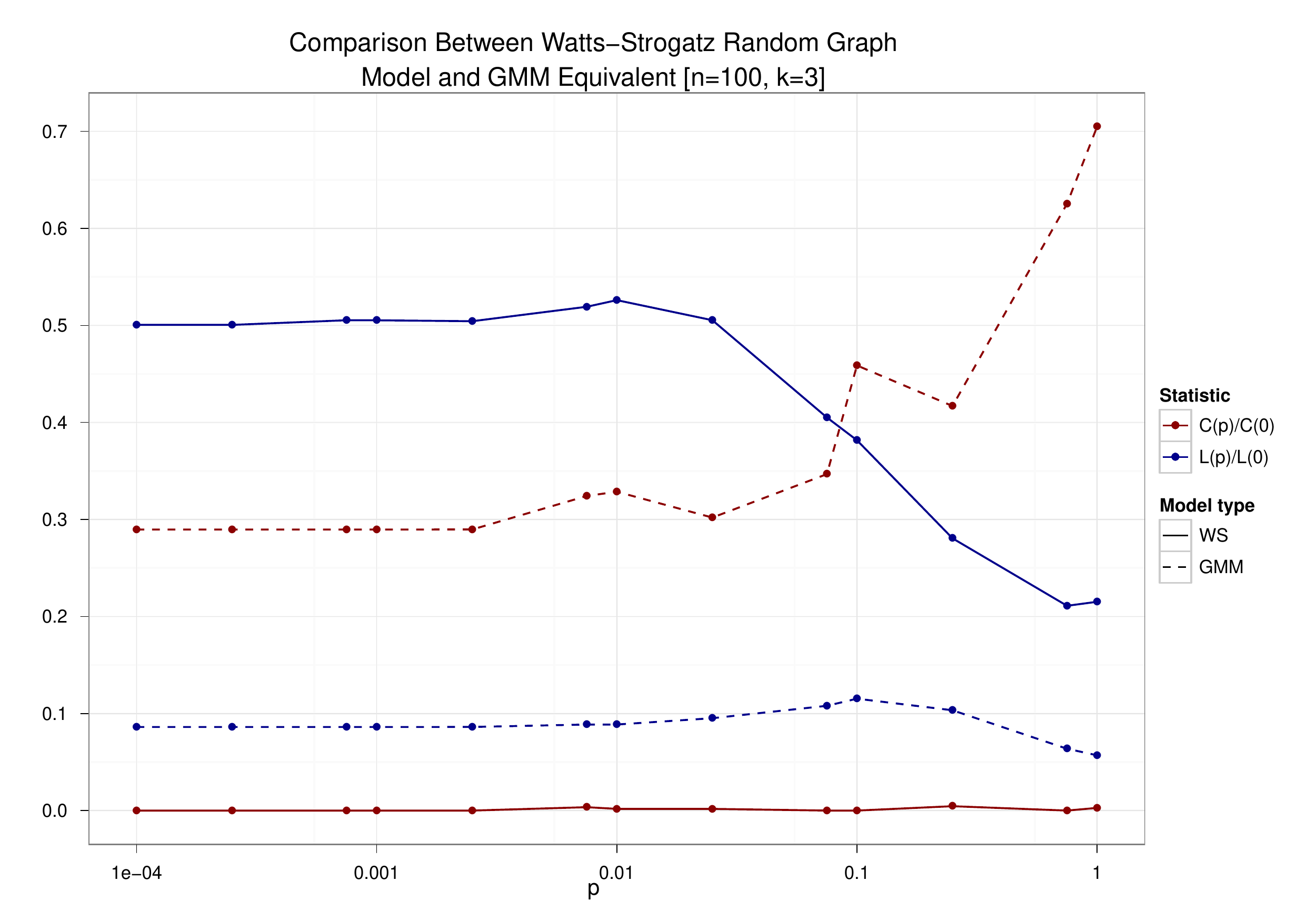}
    \caption{The above graph compare the classic Watts-Strogatz (WS) model with a GMM equivalent using both normalized clustering coefficient ($C(p)/C(0)$) and characteristic path length ($L(p)/L(0)$).  These values have been normalized by the clustering coeffecient and characteristic path length of a regular lattice of $n=100$ and $k=3$, as in the original WS paper.  Each point in the graph represents the mean value for both metrics over 20 simulations for both the classic WS and the GMM simulations.  For all simulations $n=100$ and $k=3$.  The y-axis represents variation in the $p$ in log-scale.  This is the probability of being rewired in the classic model, and the probability of connecting to the main component in the GMM. Red points and are clustering coefficients, and blues are characteristic path length.  Solid lines connect points from the classic WS model, while dotted for the GMM simulations.  The grey bars in each graph are the densities for each degree count in the graphs.  The red dotted line is the theoretical binomial fit given the distribution.}
    \label{fig:ws-comp}
\end{figure}

In this experiment a GMM is specified that attempts to recover the WS model with the framework of motif modeling.  The original WS model does not allow for new nodes to enter the network.  It can be argued that this design poorly describes real social networks.  Regardless, the GMM specified for these experiments attempts to remain as faithful to the original design as possible by using the $k$ parameter to maintain mean degree in the base structure and $p$ as the probability of connecting to the main component.  The pseudo-code for the GMM implementation is available in Appendix A.

The results of these simulations are reported in Figure 9.  This graph follows the specification used in the original WS paper, where both $n$ and $k$ are fixed in all simulations and $p$ varies between zero and one on a log-scale.  Unlike the original paper, however, $n=100$ and $k=3$ as compared to the values of $n=10,000$ and $k=10$.  As mentioned, a 10,000 node network is very large and not necessarily representative of the small-world networks this model is meant to describe.  Likewise, initializing the model with a regular lattice with mean degree 10 assumes a lot of structure at the outset.  The smaller and sparser networks in this specification are a better benchmark for comparing the GMM simulations with the classic WS results.

The results are not as straightforward as the ER simulations, but are quite telling.  Each point in Figure 9 represents the average of that value for that network specification over 20 simulations.  Thirteen variations of $p$ were used in this experiment; therefore, the figure represents the results of 260 simulations.  The solid lines correspond to the classic WS model, and dashes the GMM simulations.  Red dots represent mean values for the normalized clustering coeffecient.  Blue dots the mean characteristic path length.

There are two striking observations from this data.  First, the classic WS model overall performs poorly at these small scales.  The clustering coefficients in the WS model are flatlined at zero, almost regardless of the value of $p$.  Moreover, for very low $p$ the average shortest path length in the original WS model remains relatively high until $p>0.01$.  Alternatively, the GMM models are performing much better at generating networks with small world structure.  

In fact, in all cases the GMM simulations outperform the classic WS model in this experiment.  The clustering coefficients are higher for all simulations, and gradually increase as $p$ increases.  Unlike the WS model, wherein the clustering coeffecient is flat at zero.  Interestingly, the average shortest path length in the GMM simulations is low and only slightly affected by increases in $p$.  One of the reasons that the GMM results may be performing better is the framework is a much more natural way to model a small world structure.  Rather than assuming a fixed initial structure and randomly permuting edges, the GMM is effectively bringing in clustered groups through the motif.  The motifs create high localized structure, which in turn decreases the average shortest path for networks generated this way.

As before, a key strength of the GMM modeling framework is its flexibility.  With the ability to specify tailored growth rules GMM can describe a very robust set of network evolutionary mechanisms.  In this case, using graph motifs to model growth is a much more intuitive way of generating small world networks.  In the third and final experiment a GMM is specified which attempts to recover the classic Barab\'{a}si-Albert (BA) preferential attachment model.

The BA model is motivated by the strong empirical observations of heavy-tailed degree distributions in large networks.  This comes from very few nodes acting as massive hubs in large networks.  The vast majority of nodes have few connections, while a minority have very many.  Networks with these features are often referred to as ``power-law networks,'' because their degree distributions are so heavily skewed.  Some examples of networks that have been shown to exhibit these features are the World Wide Web \citep{albert_1999}, protein interaction networks \citep{Ito_2000}, email networks \citep{Ebel_2002}, and country-to-country war dyads \citep{Roberts_1998}.

In these networks nodes that have many edges tend to get more new edges faster than those with fewer edges.  This has been described as a ``rich get richer'' dynamic, or preferential attachment.  These nodes have some inherent attribute that makes them more likely to get new connections.  In the context of the World Wide Web Google is a massive hub, primarily because other web sites want to connect to Google in order to be found.  In war, some countries are much more aggressive than others and tend to get into more militarized disputes.  This dynamic is the center piece of the BA model.  

The classic model takes two parameters: $n$ and $m$.  The first, $n$, is simply the number of nodes in the final graph.  The $m$ parameter is how many edges each new node will make when entering the network.  The model always begins with a simple base structure, usually a three node line graph.  Then, nodes iteratively enter the network forming $m$ ties to the nodes in the base graph as a function of the number of edges each node in the base graph already has.  That is, nodes with more edges have a higher probability of forming ties with new nodes, i.e., preferential attachment based on degree.  With this framework the BA model produces networks with heavy-tailed degree distributions.

\begin{landscape}
    \begin{figure}[H]
        \centering
        \subfigure[$n=100$, $m=1$]{\label{fig:BA1}\includegraphics[width=.55\textwidth]{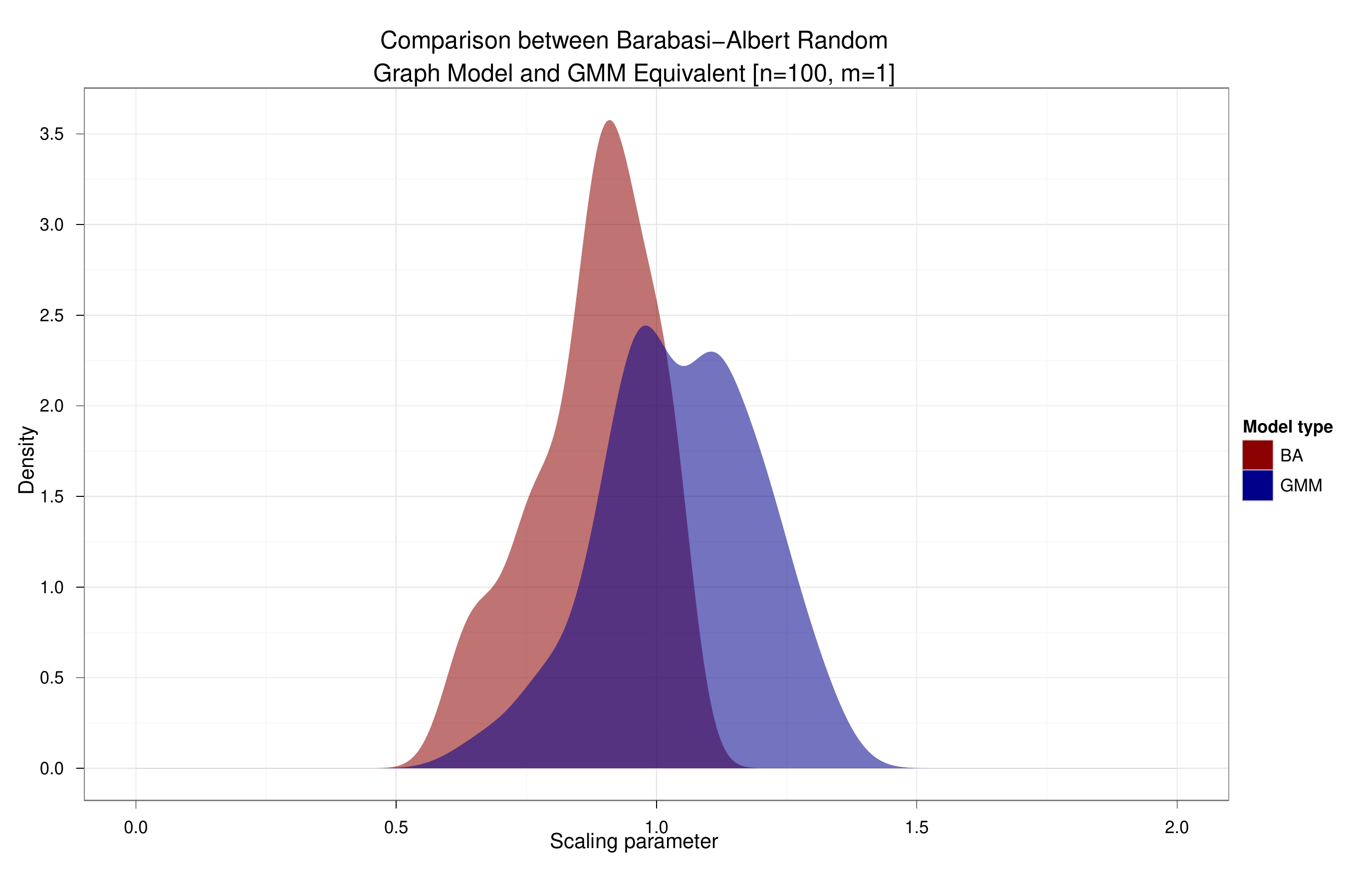}}
        \subfigure[$n=100$, $m=3$]{\label{fig:BA3}\includegraphics[width=.55\textwidth]{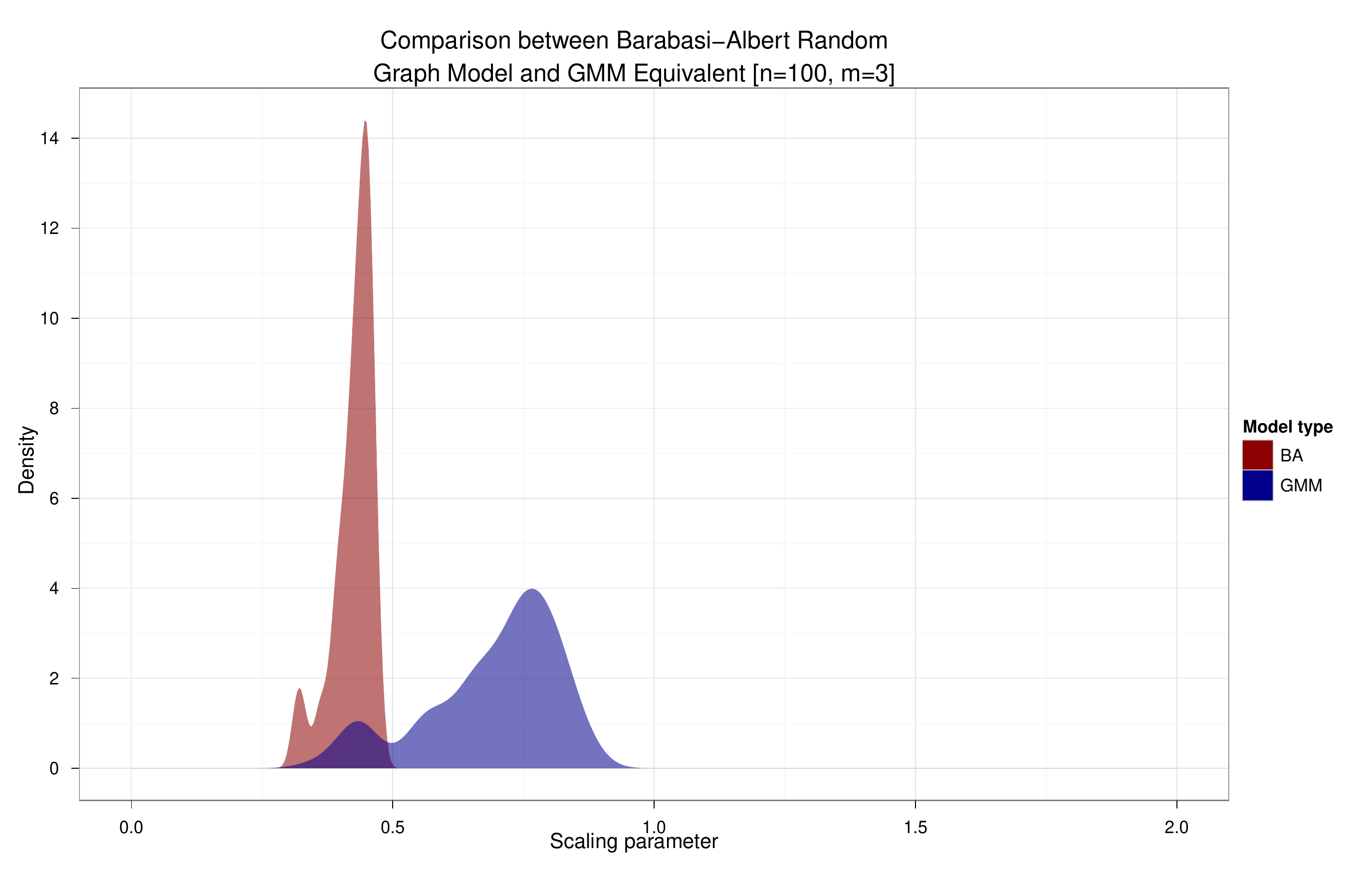}}
        \subfigure[$n=100$, $m=5$]{\label{fig:BA3}\includegraphics[width=.55\textwidth]{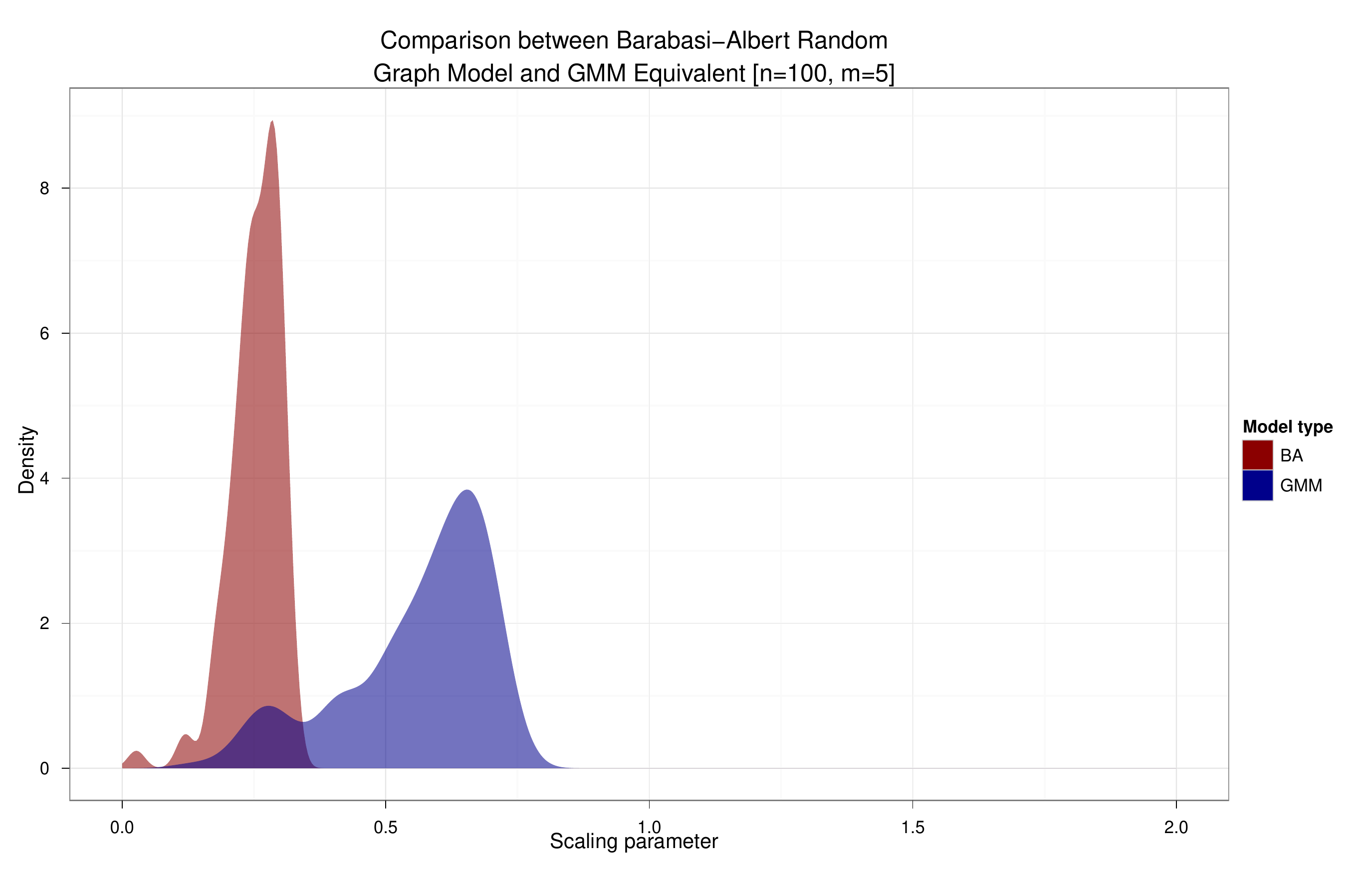}}
        \subfigure[$n=100$, $m=7$]{\label{fig:BA3}\includegraphics[width=.55\textwidth]{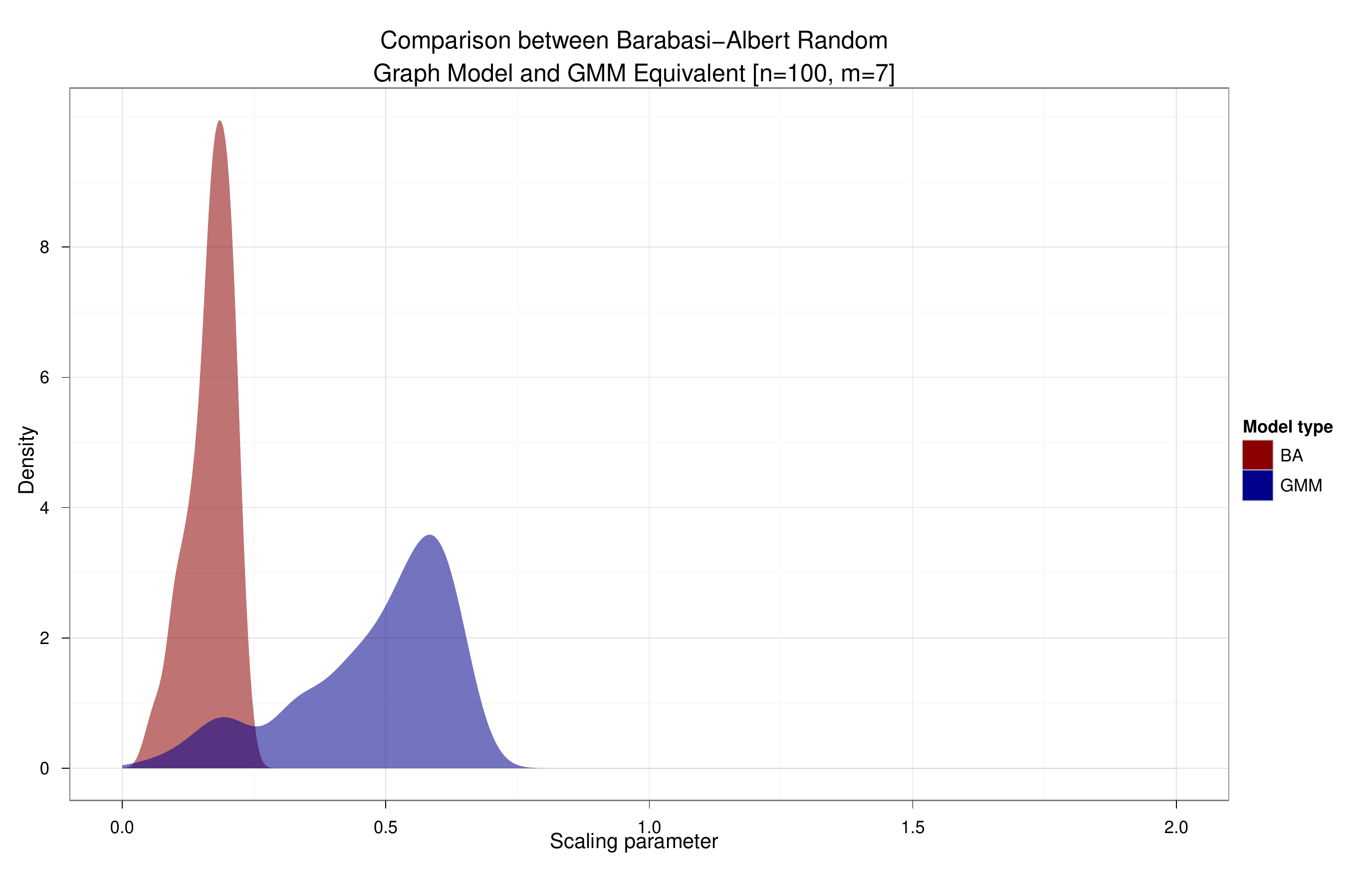}}
        \caption{The above graphs compare kernel density plots of the power-law scaling parameters calculated from the degree distributions for networks generated with a classic Barab\'{a}si-Albert (BA) random graphs networks with those from a GMM equivalent.  In all simulations $n=100$, and starting in the upper-left panel $m=1$.  Moving counter clockwise through the panels $m$ increases to 3, 5, and 7.  These parameters were estimated graphically.  In each graph the red curve represents the density of the scaling parameters for 100 simulations of the class BA model for that parameterization.  The blue curves are the cumulative results of GMM simulations with the same specifications, but with varying base sizes.  Each blue curve contains results from 25 GMM simulations using BA models as the base with increasing size, $n={20,40,60,80}$.}
        \label{fig:ba-comp}
    \end{figure}
\end{landscape}

\begin{landscape}
    \begin{figure}[H]
        \centering
        \subfigure[$n=100$, $m=1$]{\label{fig:BA1}\includegraphics[width=.55\textwidth]{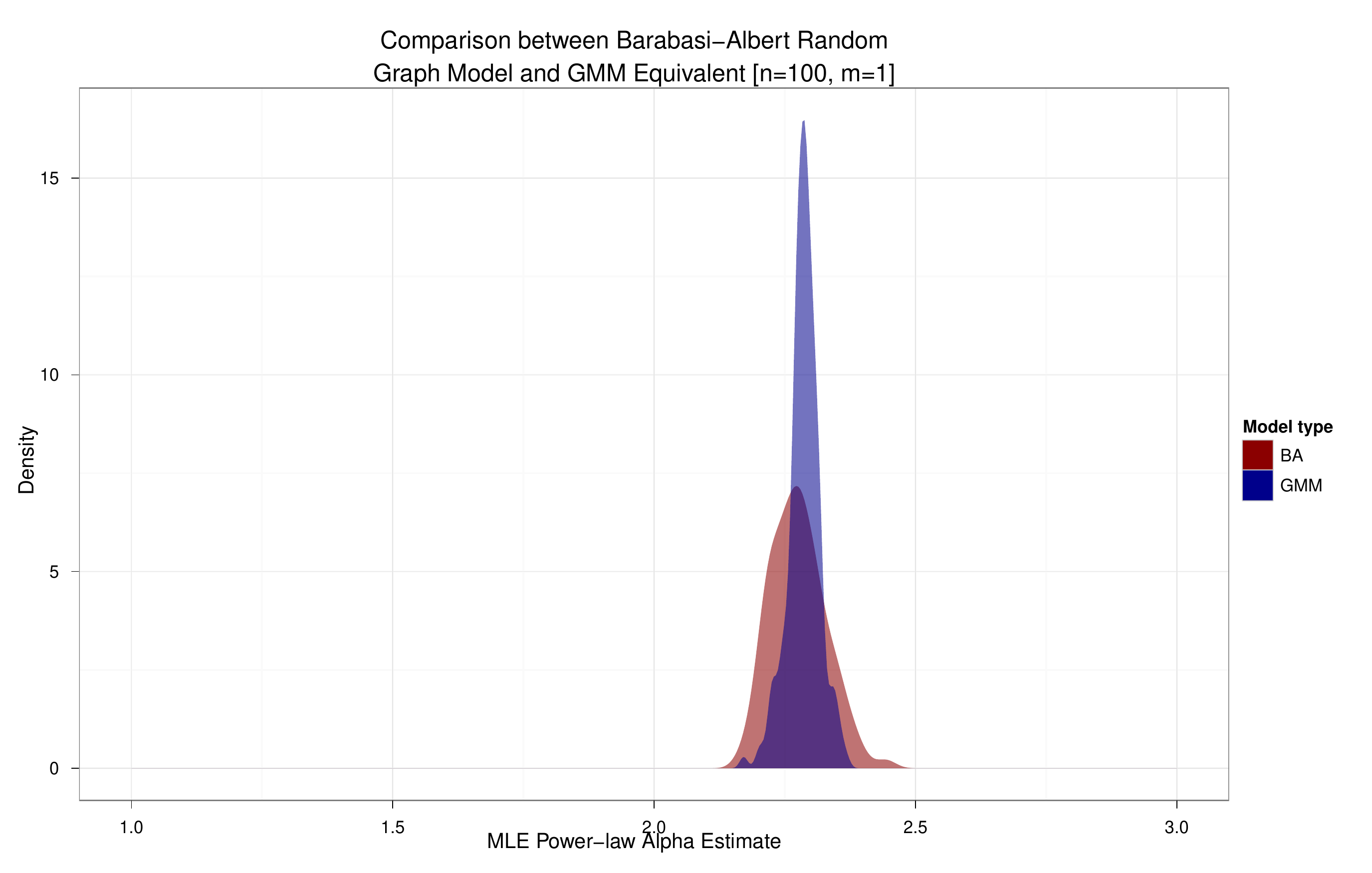}}
        \subfigure[$n=100$, $m=3$]{\label{fig:BA3}\includegraphics[width=.55\textwidth]{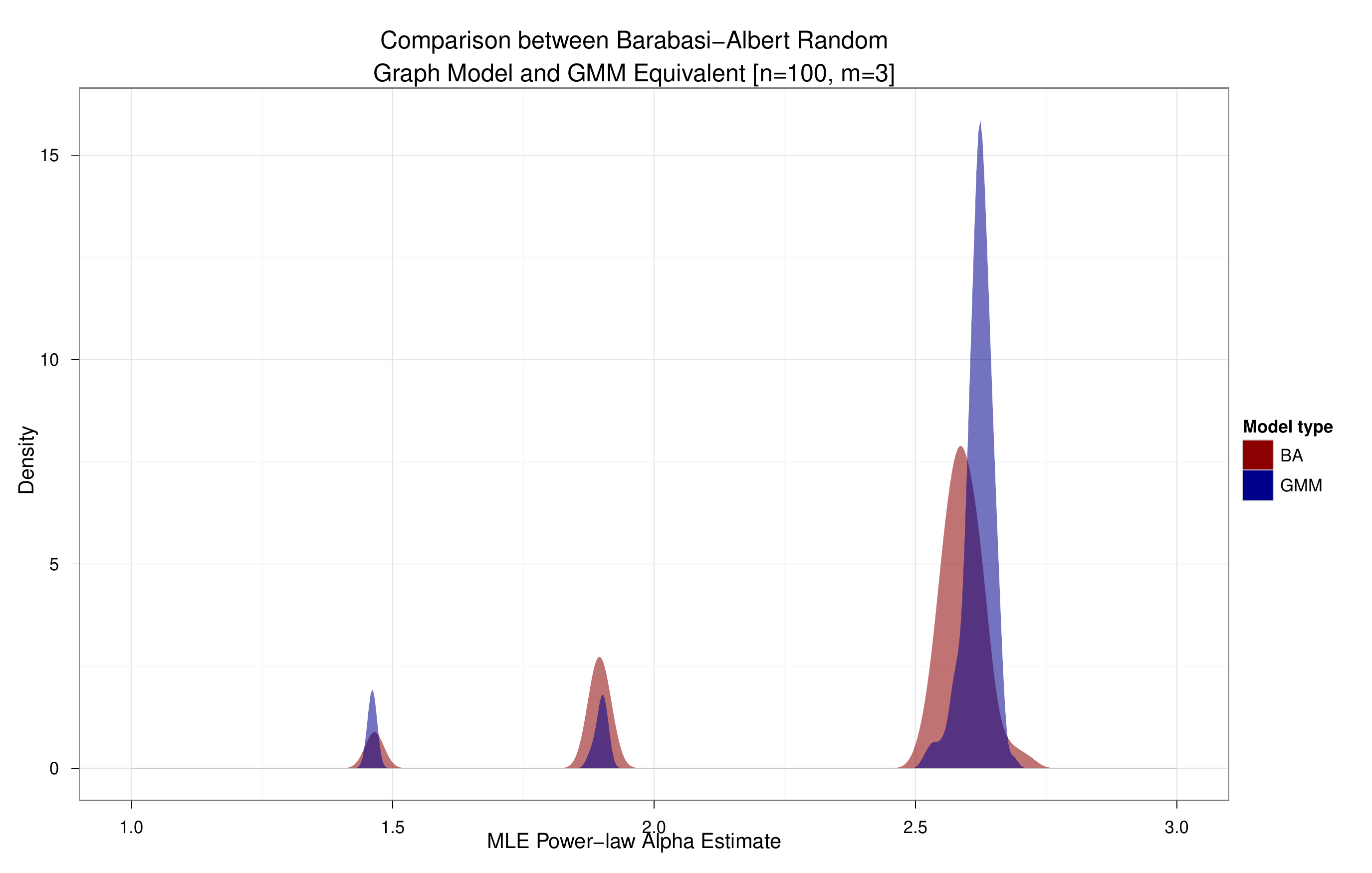}}
        \subfigure[$n=100$, $m=5$]{\label{fig:BA3}\includegraphics[width=.55\textwidth]{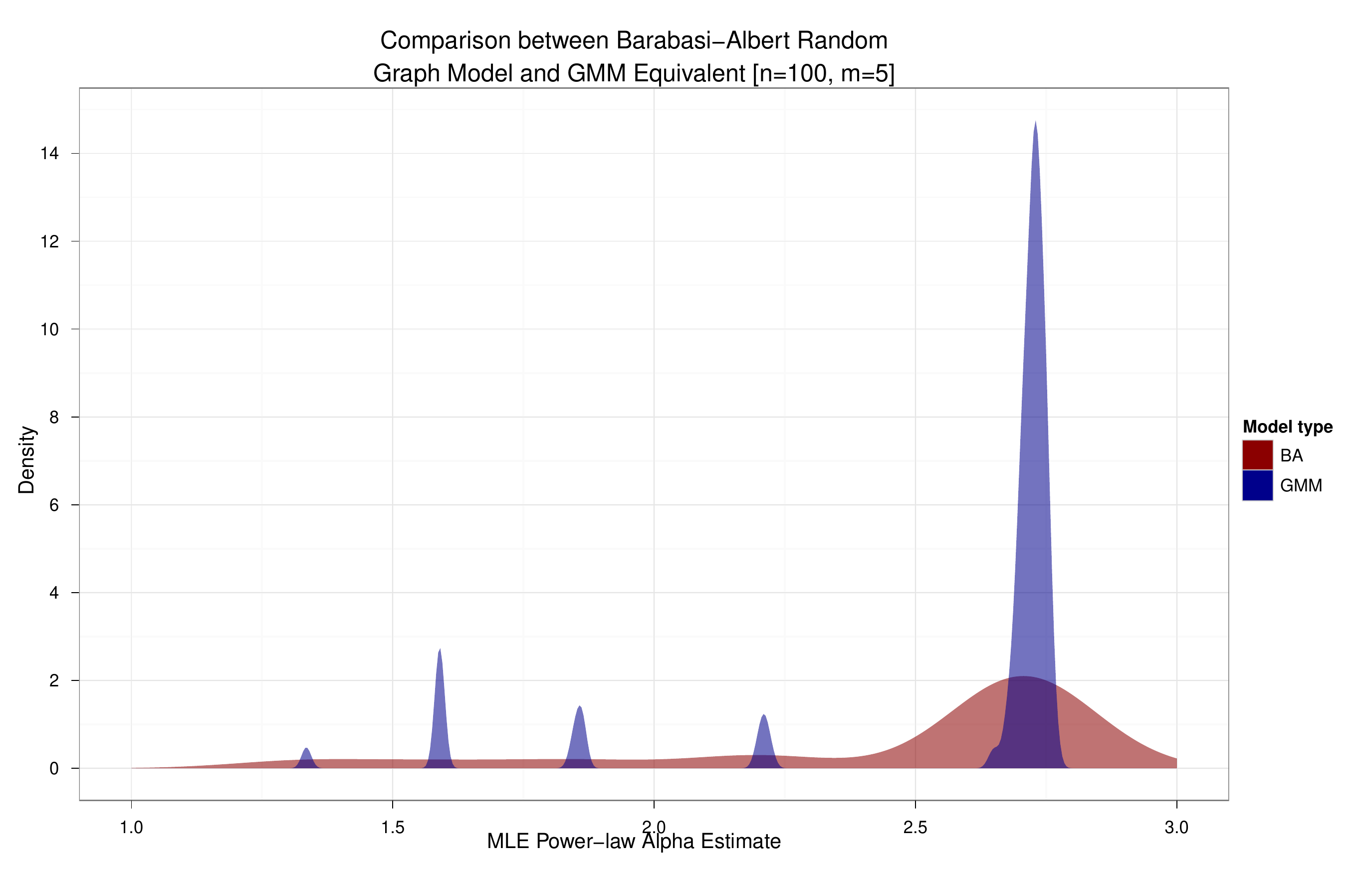}}
        \subfigure[$n=100$, $m=7$]{\label{fig:BA3}\includegraphics[width=.55\textwidth]{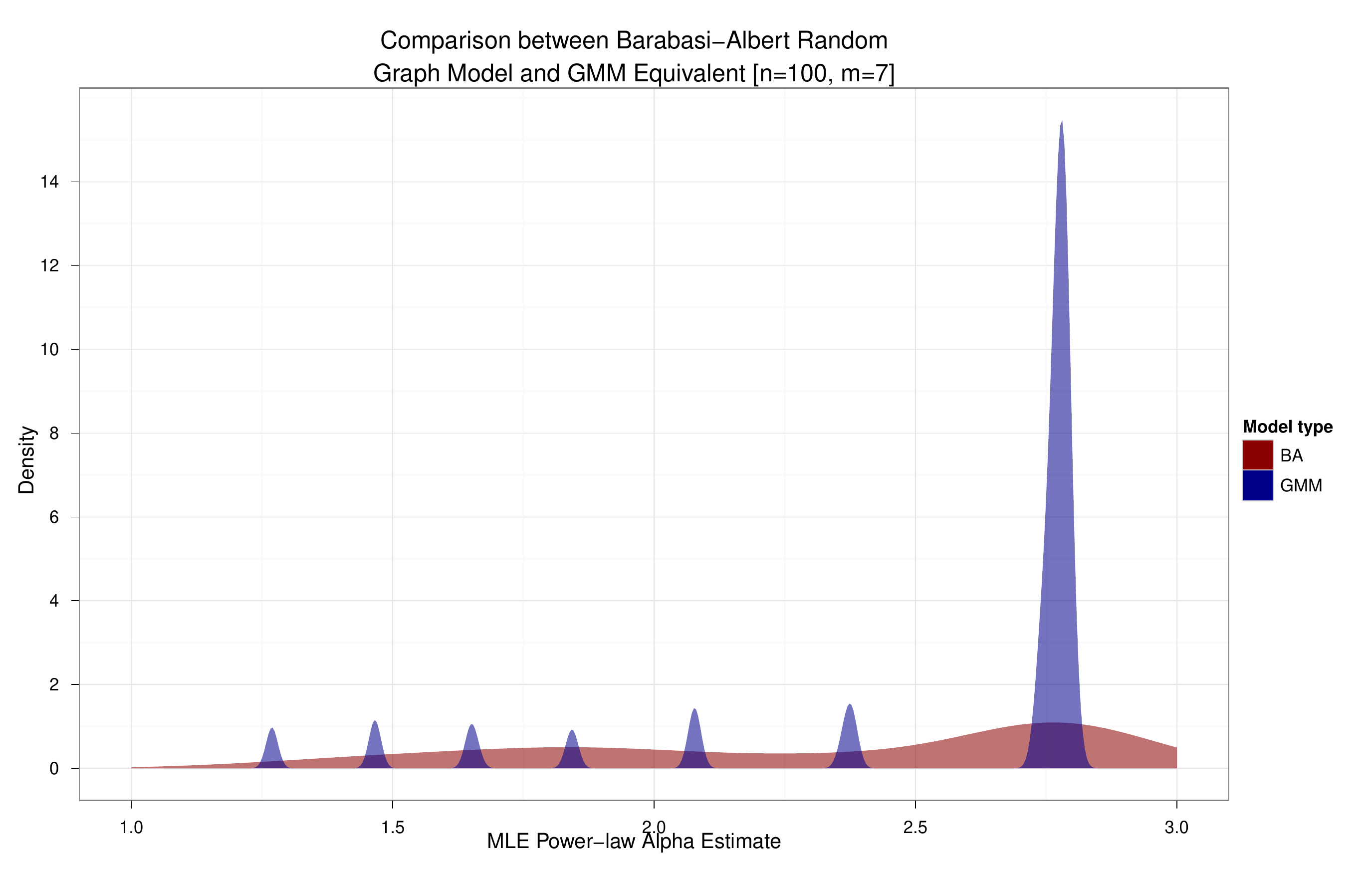}}
        \caption{The above graphs compare kernel density plots of the alpha parameters for fits to a power-law distribution from degree distributions for networks generated with a classic Barab\'{a}si-Albert (BA) random graphs networks and those from a GMM equivalent.  In all simulations $n=100$, and starting in the upper-left panel $m=1$.  Moving counter clockwise through the panels $m$ increases to 3, 5, and 7.  Unlike in Figure 10, where the scaling parameter is estimated graphically, the alpha values here are estimated using maximum likelihood.  In each graph the red curve represents the density of the scaling parameters for 100 simulations of the class BA model for that parameterization.  Curves match Figue 10.}
        \label{fig:ba-comp}
    \end{figure}
\end{landscape}

To specify a GMM that recovered the BA model is simple.  Using a small network generated with the classic BA model the GMM takes an $m$ parameter and connects the motif $m$ times to the base structure as a function of the nodal degree in the base structure.  As in the WS experiment the size of the networks in both the BA and GMM are held constant at $n=100$.  Variation is done on the $m$ parameter, starting with  $m=1$ and increasing to 3, 5, and 7.  With the classic BA models, 100 simulations were generated for each parameterization.  For the GMM 25 simulations were conducted using BA models as the base with increasing size, $n={20,40,60,80}$.  This variation was included to investigate if the quality of the GMM results were a function of the size of the base structure.  A pseudo-code implementation of the growth rule used in the GMM is included in Appendix A.

The ability of these models to produce power-law-like networks is tested by fitting the resulting degree distributions to a power-law.  A common method for fitting these degree distributions is to do so graphically.  This is done by fitting a line to the normalized degree distributions in log-scale.  The slope of that line is an estimate of the power-law scaling parameter $\alpha$.  Using this method an estimate of $\alpha>2$ is considered a ``good fit'' to a power-law.  This is the method used in the original BA piece, and the results of this test on the experimental networks are reported in Figure 10.

To compare the results from the BA model and the GMM simulations the scaling parameters estimated for all graphs are grouped together and kernel density plots are used to illustrate the frequency of scaling parameters.  The red curves in Figure 10 represent the densities of scaling parameters from 100 BA simulations, and the blue curves the 100 GMM simulations with varying base graph sizes.  Starting in the upper-left panel $m=1$, and  moving  clockwise through the panels $m$ increases to 3, 5, and 7.  The results of these tests highlight both strengths and weaknesses of the GMM framework in this context.

Panel (a) of Figure 10 shows that the GMM produces very similar scaling parameters to the BA model.  In this case $m=1$, so only one node from the motif is being connected to the base structure during the GMM simulation.  It is also clear that even in panel (a) the GMM simulation produces more variation in the scaling parameter.  Moving through the other panels, the contrast is much starker.  The GMM is producing more varied and less similar scaling parameters than the classic BA model.  The peaks in the blue curve seem to indicate that variation in the size of the base structure for the GMM does affect the results.  Further numeric investigation of these results confirm this.  What is interesting, however, is that despite this variation the GMM simulations are producing better results in terms of power-law fit.

When directly comparing the results of the scaling parameters using the graphical estimation technique the GMM does better.  But, considering $\alpha>2$ as the rule of thumb, in fact neither model is producing good results.  This may be because of the relatively small sizes of the simulated networks.  At this scale the GMM is on average producing networks that have a better fit by this metric.  Using a graphical estimate of a power-law fit, however, is undesirable.  It has been shown that graphical estimates, like the one used in the classic BA piece and reported in Figure 10, are known to produce systematically biased estimates \citep{Clauset_2009}.  A more precise method is to use maximum-likelihood to estimate the scaling parameter.  Such a method for fitting power-laws has been proposed \citep{Newman_2005}, and is used to re-estimate the scaling parameters for all of the networks in Figure 11.

The most immediate difference between the results of the graphical estimates and the maximum likelihood is the considerable reduction in variance of the scaling parameter in the GMM simulations.  Moreover, estimates using the MLE method significantly increase the variance for scaling parameters from BA networks.  Unlike the results from the graphical estimates, in this case the GMM simulations are producing networks with very similar scaling parameters.  This can be seen by the the tight overlaps on the curves in all of the panels in Figure 11. Finally, these re-estimates reinforce the observation that the size of the base structure affects results in the GMM.  This is particularly noticeable in panels (c) and (d), where there are several small curves for the GMM simulations.

The results of these experiments are very encouraging.  As a proof of concept exercise, the GMM used here were all able to successfully recover the data generating processes described by these three classic models.  In addition, in many cases the GMM were able to produce networks that better achieved the structural features described by the classic models.  These results also reveal a potential weakness of the GMM.  The final experiment highlights the sensitivity of the GMM to the size of the base structure.  


\section*{Conclusions} 
\label{sec:conclusions}

In this paper I have introduced an alternative technique for modeling network evolution using graph motifs, i.e the GMM.  This method differs greatly from current models in its core assumptions.  First, the GMM framework assumes networks evolve through the addition of nodes with exogenous structure.  When new actors enter a network they do so with some amount of preexisting structure.  This structure should therefore be used to model network growth.  Second, future structure in a network will resemble current structure.  This assumption relies on the observation that networks exhibit considerable fractal scaling as they increase in complexity.  To estimate this the GMM uses counts subgraph isomorphisms of a set of graph motifs to measure the frequency of various network structures within a given graph.  Using these assumptions, the GMM is constructed as computational framework for simulating network growth.

The basic GMM framework has been implemented as the \texttt{GMM} package in the \texttt{Python} programming language.  Relying on high-quality scientific computing packages already available in \texttt{Python}, this package allows for the specification of a near boundless set of GMM to model any number of networks.  To test this framework various GMM are specified that attempt to recover three classic random graph models: the Erd\H{o}s-R\`{e}nyi binomial random graph, the Watts-Strogatz ``small world'' model, and the Barab\'{a}si-Albert preferential attachment model.  In each case the GMM  is able to not only successfully recover these models, but also often outperforms them.  These results are very encouraging for the use of GMM in modeling many different types of network growth dynamics.  

This work has many potential contributions to political science.  As stated, much of the data studied in the social sciences can be modeled as a network.  More specifically, this data often represents relationships among people.  While there have been great advances in techniques for modeling these relationships, current methods lack a flexible framework for modeling network evolution.  The dynamics of human interactions are both complex and subtle.  Attempting to force these complexities into overly simple models massively limits the types of networks that can be studied.  By using a more flexible framework, such as the GMM proposed here, social science researchers may be able to specify models that capture these elusive dynamics.  Experiments like the ones described above can be used to explore the ramifications of these dynamics, and how they affect social outcomes.

It is important to note, however, that the technique proposed here also has many limitations.  As was drawn out in the experiments, the results of GMM appear sensitive to the size and structure of the initial base graph.  Additional tests must be conducted to determine the nature of this sensitivity.  Also, in its current form the growth and termination rules are treated as fixed for the entire simulation.  Conceptually, this is useful because it simplifies the construction of a model and provides a basis for interpretation.  In some cases, however, it may be useful to ``endogenize'' these rules given the initial state of the base structure.  Consider the case where the base structure is unknown to the modeler at the outset.  Here, we may want rules that emerge as the result of this structure given the context of our modeling task.  In this case endogenous rule generation will be necessary.  Furthermore, the notion of ``learning rules'' is a potential extension.  Growth rules could be tuned given the evolution of the network through the iterations of the model.  These adaptations, however, make model interpretation more difficult.

Finally, a more general observation is the GMM is poorly suited to model non-human networks.  There are many networks for which exogenous growth may be a contradiction, such as physical networks like transportation or telecommunication.  Also, the evolution of some biological networks may be poorly modeled using graph motifs, such as protein-interaction or neural networks.  In cases where network evolution via graph motifs is not feasible then the GMM should not be applied.  That said, the primary motivation for this work is to attempt to establish a framework for modeling the evolution of human social networks.

Beyond these theoretical limitations, there are also some technical limitations.  A linchpin of the model is the need to count subgraph isomorphism in order to form beliefs about future network structure.  As stated, this problem is known to be NP-complete, and therefore the method scales very poorly as either $\tau$ or the complexity of the base structure increases.  In practice both of these model parameters must be relatively small to compute results in a reasonable amount of time.  Improving the speed of the VF-2 algorithm is a computer science problem, and therefore beyond the scope of this research.  With current technology, however, there are methods for improving runtime as the networks scale.  First, rather than recomputing the probability distribution at every iteration this could remain static, meaning that subgraph isomorphism would only need to calculated once.  Additionally, these counts could easily be done in parallel in a high-performance computing environment.  Future version of the \texttt{GMM} package will allow for such distributed computing.

Additional improvements need to be made to future versions of the software.  Better accounting for simulation statistics, including runtime, growth metrics, probability mass convergence and iterative changes to the base network.  For example, given the potential path dependence of the model from the base structure it would be very useful to have some knowledge of the distribution of motifs used in a given simulation.  This information could be used to compare and interpret results from multiple runs of the same model.  Finally, utilizing the ERGM literature, considerations for the quality of ``model fit'' within the context of the GMM must be made.  A large advantage of ERGM models is the ability to compare model fitness.  In order to more fully understand how these two modeling techniques differ direct comparisons must be made across a large class of networks.  This type of research, therefore, will constitute a large portion of the future effort in the work.


\bibliographystyle{chicago}
\newpage
\setstretch{1.0}
\bibliography{gmm}
\newpage

\setstretch{1.5}

\newpage
\section*{Appendix A: Growth rules for experiments} 
\label{sec:appendix_c_growth_and_termination_}

The following are the \texttt{Python} functions used for the growth and termination rules in the GMM specified to generate the simulated network described in the \emph{Recovering classic models} section above.  

\begin{algorithm}[H]
    \begin{algorithmic}
        \REQUIRE $G,H,p$
        \STATE $G=G[H]$
        \COMMENT{Compose $H$ with $G$} 
        \FOR{$i$ in $H$}
            \FOR{$j$ in $G$}
                \STATE $ran=RANDOM(low=0, high=1)$
                \COMMENT{Draw a random value from a uniform distribution}
                \IF {$ran <= p$}
                    \STATE $G=EDGE(G,i,j)$
                    \COMMENT{If random draw less than $p$, create edge}
                \ENDIF
                \ENDFOR
                \ENDFOR
                \COMMENT{For each node in $H$ test if it will connect to each node in $G$}
        \RETURN $G$
    \end{algorithmic}
    \caption{Pseudo-code growth rule for Erd\H{o}s-R\`{e}nyi binomial random graph}
\end{algorithm}

\begin{algorithm}[H]
    \begin{algorithmic}
        \REQUIRE $G,H,k,p$
        \STATE $G=G[H]$
        \COMMENT{Compose $H$ with $G$}
        \STATE $SHUFFLE(H)$
        \COMMENT{Shuffle the nodes in $H$}
        \FOR{$i$ in $H[0:k]$}
            \FOR{$j$ in $G$}
                \STATE $ran=RANDOM(low=0, high=1)$
                \COMMENT{Draw a random value from a uniform distribution}
                \IF {$ran <= p$}
                    \STATE $G=EDGE(G,i,j)$
                    \COMMENT{If random draw less than $p$, create edge}
                \ENDIF
                \ENDFOR
                \ENDFOR
                \COMMENT{With probability $p$, connect $k$ nodes from $H$ to all nodes in $G$.}
        \STATE $FULL(G)$
        \COMMENT{Ensure that $G$ is fully connected.}
        \RETURN $G$
    \end{algorithmic}
    \caption{Pseudo-code growth rule for Watts-Strogatz ``small world'' model}
\end{algorithm}

\begin{algorithm}[H]
    \begin{algorithmic}
        \REQUIRE $G,H,m$
        \STATE $G=G[H]$
        \COMMENT{Compose $H$ with $G$}
        \STATE $DEG=DEGREE(G)$
        \COMMENT{Calculate degree of nodes in $G$}
        \FOR{$i$ in $m$}
            \STATE $EDGE\_MADE=FALSE$
            \WHILE{$EDGE\_MADE==FALSE$}
                \STATE $p=RANDOM(low=0, high=1)$
                \COMMENT{Draw a random value from a uniform distribution}
                \STATE $j=RAN\_NODE(G)$
                \COMMENT {Pick a random node in $G$}
                \IF {$p <= DEG[j]$}
                    \STATE $G=EDGE(G,i,j)$
                    \COMMENT{If random draw less than degree of $j$, create edge}
                \ENDIF
                \ENDWHILE
                \ENDFOR
                \COMMENT{Connect $m$ nodes from $G$ to $H$ as function of degree in $G$}
        \RETURN $G$
    \end{algorithmic}
    \caption{Pseudo-code growth rule for Barab\'{a}si-Albert preferential attachment model}
\end{algorithm}


\end{document}